\documentclass[12pt]{article}
\pdfoutput=1
\usepackage{amssymb, amsmath,amsfonts}
\usepackage{multirow}
\usepackage{mathrsfs}
\usepackage{array}
\usepackage{cite}
\usepackage{booktabs}
\usepackage{tikz}
\usepackage{pgfplots}
\usepackage{float}
\usepackage{subfigure, graphicx}
\usepackage[pdftex, bookmarks=true,colorlinks,linkcolor=red,urlcolor=blue,citecolor=blue]{hyperref}
\usepackage{cite}
\usepackage{slashed}
\usepackage{tabularx}

\makeatletter\@addtoreset{equation}{section}\makeatother

\def\be{\begin{equation}}
\def\ee{\end{equation}}
\def\bea{\begin{eqnarray}}
\def\eea{\end{eqnarray}}

\def\Dslash{\,\,{\raise.15ex\hbox{/}\mkern-12mu D}}
\def\Dbarslash{\,\,{\raise.15ex\hbox{/}\mkern-12mu {\bar D}}}
\def\delslash{\,\,{\raise.15ex\hbox{/}\mkern-9mu \partial}}
\def\delbarslash{\,\,{\raise.15ex\hbox{/}\mkern-9mu {\bar\partial}}}
\def\pslash{\,\,{\raise.15ex\hbox{/}\mkern-9mu p}}
\def\calDslash{\,\,{\raise.15ex\hbox{/}\mkern-12mu {\cal D}}}

\makeatletter\@addtoreset{equation}{section}\makeatother

\newcommand{\preprint}[1]{\begin{table}[t]  
             \begin{flushright}               
             {#1}                             
             \end{flushright}                 
             \end{table}}                     
\renewcommand{\title}[1]{\vbox{\center\LARGE{#1}}\vspace{5mm}}
\renewcommand{\author}[1]{\vbox{\center#1}\vspace{5mm}}
\newcommand{\address}[1]{\vbox{\center\em#1}}

\def\arXiv#1{\href{http://arxiv.org/abs/#1}{arXiv:#1}}
\def\arXiv#1#2{\href{http://arxiv.org/abs/#1}{arXiv:#1}}

\def\doi#1#2{\href{http://doi.org/#1}{#2}}

\begin{document}
\unitlength = .8mm

\begin{titlepage}
\vspace{.5cm}
\preprint{}

\date{\today}
\begin{center}
\hfill \\
\hfill \\
\vskip 1cm

\title{\boldmath 
Interior of helical black holes}
\vskip 0.5cm
{Yan Liu$^{\,a,b}$}\footnote{Email: {\tt yanliu@buaa.edu.cn}} and
{Hong-Da Lyu$^{\,a}$}\footnote{Email: {\tt hongdalyu@buaa.edu.cn}}

\address{${}^a$Center for Gravitational Physics, Department of Space Science, \\ and International Research Institute
of Multidisciplinary Science,
\\ Beihang University,  Beijing 100191, China}
\address{${}^{b}$Peng Huanwu Collaborative Center for Research and Education, \\Beihang University, Beijing 100191, China}

\end{center}
\vskip 1.5cm

\abstract{
We study the interior structure of five dimensional neutral helical black holes in Einstein gravity and charged helical black holes in Einstein-Maxwell gravity. Inside the neutral helical black holes, the systems evolve to a stable spacelike Kasner singularity. The metric field related to the helical deformation strength exhibits oscillation behavior close to the horizon at low temperature and small helical deformation strength. Inside the charged helical black holes, we show that the inner Cauchy horizon can not exist. The systems also evolve from the horizon to a stable Kasner singularity.  We find that the oscillations can exist and  there is a special feature that the oscillations occur near the horizon and before the collapse of the Einstein-Rosen bridge for the charged helical black holes. 
}

\end{titlepage}

\begingroup 
\hypersetup{linkcolor=black}
\tableofcontents
\endgroup
\section{Introduction}

Holography has established deep connections between black holes and quantum many-body systems. During the last two decades, motivated by the problems in quantum many-body physics, lots of interesting black hole solutions have been constructed, such as hairy black holes which are  dual to strongly correlated superconductors \cite{Hartnoll:2008kx}, asymptotic Lifshitz black holes which are dual to quantum critical states with Lifshitz scaling  \cite{Kachru:2008yh,Taylor:2015glc}, inhomogeneous black holes which are dual to systems with translational symmetry breaking \cite{Horowitz:2012ky} etc. Suggestive insights to the quantum many-body problems have been provided from 
black hole physics, see e.g. \cite{Blake:2022uyo} for a recent summary. 

Most of the studies on the gravitational side have been focusing on the exterior dynamics of black holes to make connections to the dual field theory. This is because of the fact that at classical level the interior dynamics is causally disconnected with the physics outside. However, the interior physics is believed to be crucial for understanding the whole dynamics of the dual quantum many-body system,  for example, the island formula for calculating the Page curve involves the interior dynamics \cite{Almheiri:2020cfm}. Therefore it is important to study the interior structure of black holes, in order to investigate the fundamental relationship between the interior dynamics and the dual quantum many-body physics.  

Recently, the rich interior dynamics for black holes in applied holography have been actively explored in  \cite{Frenkel:2020ysx,  Hartnoll:2020rwq, Hartnoll:2020fhc,Cai:2020wrp, Wang:2020nkd, Grandi:2021ajl, Mansoori:2021wxf, Liu:2021hap,Dias:2021afz, Sword:2021pfm,Cai:2021obq}.  For neutral black holes deformed by constant scalar sources in the dual theory, their interior geometries evolve smoothly from the horizon to Kasner singularities at late interior time  \cite{Frenkel:2020ysx}.\footnote{Similar behavior was also found for a large class of neutral black holes \cite{Wang:2020nkd, Mansoori:2021wxf,Liu:2021hap}.}
Later it was shown that the interior dynamics of charged black holes with scalar or vector hair becomes much richer. In \cite{Hartnoll:2020rwq} it was found that inside the charged black hole with neutral scalar hair, the inner horizon is destroyed by a collapse of the Einstein-Rosen bridge and the geometry smoothly evolves from the horizon to the Kasner singularity. When the scalar hair is charged, i.e. in the holographic superconductor model, close to the transition temperature the interior geometry evolves from the horizon to the singularity, which typically includes the collapse of the Einstein-Rosen bridge, Josephson oscillation and the (successive) Kasner epochs with Kasner inversion or transition \cite{Hartnoll:2020fhc}. Quite similar behaviors were also found in the generalized holographic superconductor with an axion field \cite{Sword:2021pfm} and the holographic p-wave superconductor \cite{Cai:2021obq}, while special behavior was also found, such as the absence of the Einstein-Rosen bridge collapse 
close to the would-be inner horizon for special parameters of the holographic superconductors with axion fields.

It is well known that the generic asymptotic solutions near spacelike singularities in Einstein gravity with generic initial conditions have been studied in the seminal work by Belinski, Khalatnikov and Lifshitz (BKL) \cite{BKL,BKL-book}. The asymptotic AdS black holes with scalar or vector hair have special singularity structure since there are some symmetries in the system. For these black holes the universal dynamics of Kasner epochs close to the singularity  were further analyzed based on the billiard approach \cite{Henneaux:2022ijt} and were shown to be consistent with the BKL analysis. Nevertheless, the numerical investigation of particular black holes with scalar or vector hair uncovers all the interior structure from the horizon to the singularity and might provide further fundamental connection to the dual field theory. Therefore, it is interesting to explore the interior dynamics for more explicit examples of black holes, including the ones dual to deformations with the  energy-momentum tensor. 
 
In this paper we will study the interior dynamics 
of five dimensional helical black holes.  In five dimensional helical black holes, the three dimensional spatial Euclidean symmetry is explicitly or spontaneously broken to Bianchi VII$_0$ symmetry, which are dual to spatially modulated phases. Helical black holes have been widely studied in the framework of AdS/CMT to describe  helical deformed four dimensional CFTs \cite{Donos:2014gya},  interaction-driven  
insulating phases along helical direction \cite{Donos:2012js}, helical current phases in Einstein-Maxwell-Chern-Simons theory  
\cite{Donos:2012wi}, helical superconductors with helical order \cite{Donos:2012gg} and so on. 
Here we will be interested in the interior dynamics of neutral helical black holes in Einstein gravity  \cite{Donos:2014gya} and charged helical black holes in Einstein-Maxwell theory respectively. 

Different from previous studies on the interior dynamics of the scalar or vector hairy black holes, for the helical black holes in our study the interior dynamics is induced by the metric field related to the helical deformation. For the neutral helical black holes, we will study the typical profiles of the metric fields of the helical black holes from the horizon to the spacelike singularity. Remarkably, we will show that at low temperature and small helical deformation strength there is an oscillatory regime close to the horizon and before reaching the Kasner epoch. There is no such  oscillation behavior for all the neutral black holes studied in the literature \cite{Frenkel:2020ysx,Wang:2020nkd, Mansoori:2021wxf,Liu:2021hap}.  For the charged helical black holes, we will first prove that there is no inner Cauchy horizon and then  present the typical profiles 
with oscillations from the horizon to the Kasner singularity. Intriguingly, the oscillation regime exists and it is located near the horizon while before the collapse of Einstein-Rosen bridge, in contrast to the behavior found in the literature where the oscillations occur after the collapse  \cite{Hartnoll:2020fhc}. In both these helical black holes, the Kasner geometry is stable. 

The paper is organized as follows. In Sec. \ref{sec2}, we first present the setups of neutral helical black holes in Einstein gravity and then study their interior structure. In Sec. \ref{sec3}, we study the interior structure of charged helical black holes in  Einstein-Maxwell gravity. 
We conclude and discuss open questions in Sec. \ref{sec:cd}. 
Some details of the calculations are collected in the appendices.

\section{Neutral helical black holes}
\label{sec2}

Neutral helical black holes exist in Einstein gravity with a negative cosmological constant by tuning a specific helical source for the energy-momentum tensor of the dual field theory \cite{Donos:2014gya}. The deformation is encoded in the metric field and in this sense the interior dynamics is driven by the ``deformed" metric field. We first shortly review the setup in  \cite{Donos:2014gya} where the exterior geometry of neutral helical black hole solutions were explored, and then study the interior geometry of these black holes. 

We consider five-dimensional Einstein gravity with the action  
\begin{align}\label{eq:ac1}
  S=\frac{1}{16\pi G}\int d^5x \sqrt{-g} \left(R+12\right)
\end{align}
where  the cosmological constant of AdS$_5$ is $\Lambda=-6$. The equations of motion are 
\begin{align}
  R_{ab}+ 4 g_{ab}=0\,.\label{eq:ee}
\end{align}

The ansatz of metric is
\begin{align}
\label{eq:ans1}
  ds^2=-gf^2 dt^2+\frac{dr^2}{g}+h^2 \omega_1^2+r^2(e^{2\alpha}\omega_2^2+e^{-2\alpha}\omega_3^2)\,,
\end{align}
where $g,f,h,\alpha$ are functions of radial coordinate $r$. Here $\omega_i$ are one-forms 
\begin{align}
\label{eq:oneform}
    \omega_1=dx_1\,, ~~~\omega_2=\cos{k x_1}dx_2-\sin{kx_1}dx_3\,,~~~ \omega_3=\sin{k x_1}dx_2+\cos{k x_1}dx_3\,
\end{align}
with constant wave-number $k$,  
satisfying Bianchi VII$_0$ algebra
\be
d\omega_1=0\,,~~~d\omega_2=k\omega_3\wedge\omega_1\,,~~d\omega_3=k\omega_1\wedge\omega_2\,.
\ee 
The AdS boundary is located at $r\to\infty$. The horizon is at $r=r_h$ while the singularity is at $r\to 0$. Note that the translational symmetry along $x_1$ direction is broken by the helical structure.
The system is anisotropic and the boundary value of the field $\alpha$ characterizes the helical source for the energy-momentum tensor of the dual field theory. 

The ansatz \eqref{eq:ans1} has 
Bianchi VII$_0$ symmetry and the solution describes a static neutral helical black hole. 
Note that \eqref{eq:ans1} is invariant under $k\to -k, x_1\to -x_1$ and here we will consider the cases with $k> 0$. 

Substituting \eqref{eq:ans1} into (\ref{eq:ee}) we obtain
\begin{align}
\label{eq:eomnetural}
\begin{split}
    \frac{f'}{f}+\frac{(-2rh^2(g\alpha'^2+2)+r(2k^2\sinh^2{2\alpha}-gh'^2)+(g+4r^2)hh')}{gh(rh'+2h)} &=0\,,\\
    g'+\frac{2[h^2(r^2g\alpha'^2+g-2r^2)+r^2(gh'^2-k^2\sinh^2{2\alpha})+r(g-4r^2)hh']}{rh(rh'+2h)} &=0\,,\\
    h''+\frac{4rh'}{g}-\frac{4h}{g}+\frac{h'}{r}-\frac{h'^2}{h}+\frac{2k^2 \sinh^2{2\alpha}}{gh} &=0\,,\\
    \alpha''+\frac{4r\alpha'}{g}+\frac{\alpha'}{r}-\frac{k^2 \sinh{4\alpha}}{gh^2} &= 0\,,
    \end{split}
\end{align}
where primes are derivatives with respect to $r$. We have four independent ODE's for four fields. 

The ansatz and EOM have three scaling symmetries
\begin{align}
    &r\rightarrow \lambda r\,,~~~ \{t,x_2,x_3\}\rightarrow \lambda^{-1}\{t,x_2,x_3\}\,,~~~g\rightarrow \lambda^2 g\,;\label{eq:sym1}\\
    &t\rightarrow \lambda t\,,~~~ f\rightarrow \lambda^{-1} f\,;\label{eq:sym2}\\
    &x_1 \rightarrow \lambda^{-1} x_1\,,~~~ h\rightarrow \lambda h\,,~~~ k\rightarrow \lambda k\,.\label{eq:sym3}
\end{align}
These symmetries are important for numerical calculations. 

There is a useful radially conserved quantity associate with the above symmetries, 
\begin{align}
\label{eq:neucc}
    Q=\frac{2r^4h}{f}\left(\frac{f^2g}{r^2}\right)' \,.
\end{align}
It can be used to check our numerical computations. 

The near horizon and near boundary expansions of the metric fields can be found in appendix \ref{app:A}. From these expansions and the above scaling symmetries, we find that the whole bulk geometry from the singularity to the boundary is completely determined by two\footnote{We have fixed the nontrivial dynamical scale due to the conformal anomaly which is related to the log terms for the expansions near the boundary \cite{Donos:2014gya}.} dimensionless free parameters $\alpha_b$ and $T/k$ of the boundary field theory, where $\alpha_b$ characterizes the strength of the helical deformation, $T$ is the temperature and $k$ characterizes the pitch of the helical structure.  

The thermodynamic quantities are 
\be
T=\frac{r_h f_h}{\pi f_b} \,, ~~~~s=\frac{4\pi r_h^2 h_h}{h_b}\,
\ee
where $r_h$ is the location of horizon, $f_h,h_h$ and $f_b,h_b$ are values of metric fields $f, h$ evaluating at the horizon and the boundary respectively. 
Since the metric fields $f, g, h, \alpha$ are smooth along the radial direction, one can integrate the system from boundary towards to the singularity to obtain the full bulk geometry. 

Note that the system has the AdS-Schwarzschild solution when   
$\alpha=0$
\be
g=r^2-\frac{r_h^4}{r^2}\,,~~ f=1\,,~~~ h=r\,.
\ee
When $\alpha_b\neq 0$ and $k\neq 0$, we have a neutral helical black hole solution with nontrivial profile of $\alpha$. In this sense we state that the interior dynamics of the helical black hole is induced by the metric field $\alpha$. 

The UV field theory is specified by the deformations of wave-number $k$ and the strength of the deformation $\alpha_b$. Equivalently, the dual field theory lives on the spacetime  \cite{Donos:2014gya}
\be
ds^2=-dt^2+\omega_1^2+e^{2\alpha_b} \omega_2^2+e^{-2\alpha_b} \omega_3^2\,
\ee
at finite temperature with $\omega_i$ defined in  \eqref{eq:oneform}.

\subsection{The interior structure}

\begin{figure}[h!]
\begin{center}
\includegraphics[
width=0.32\textwidth]{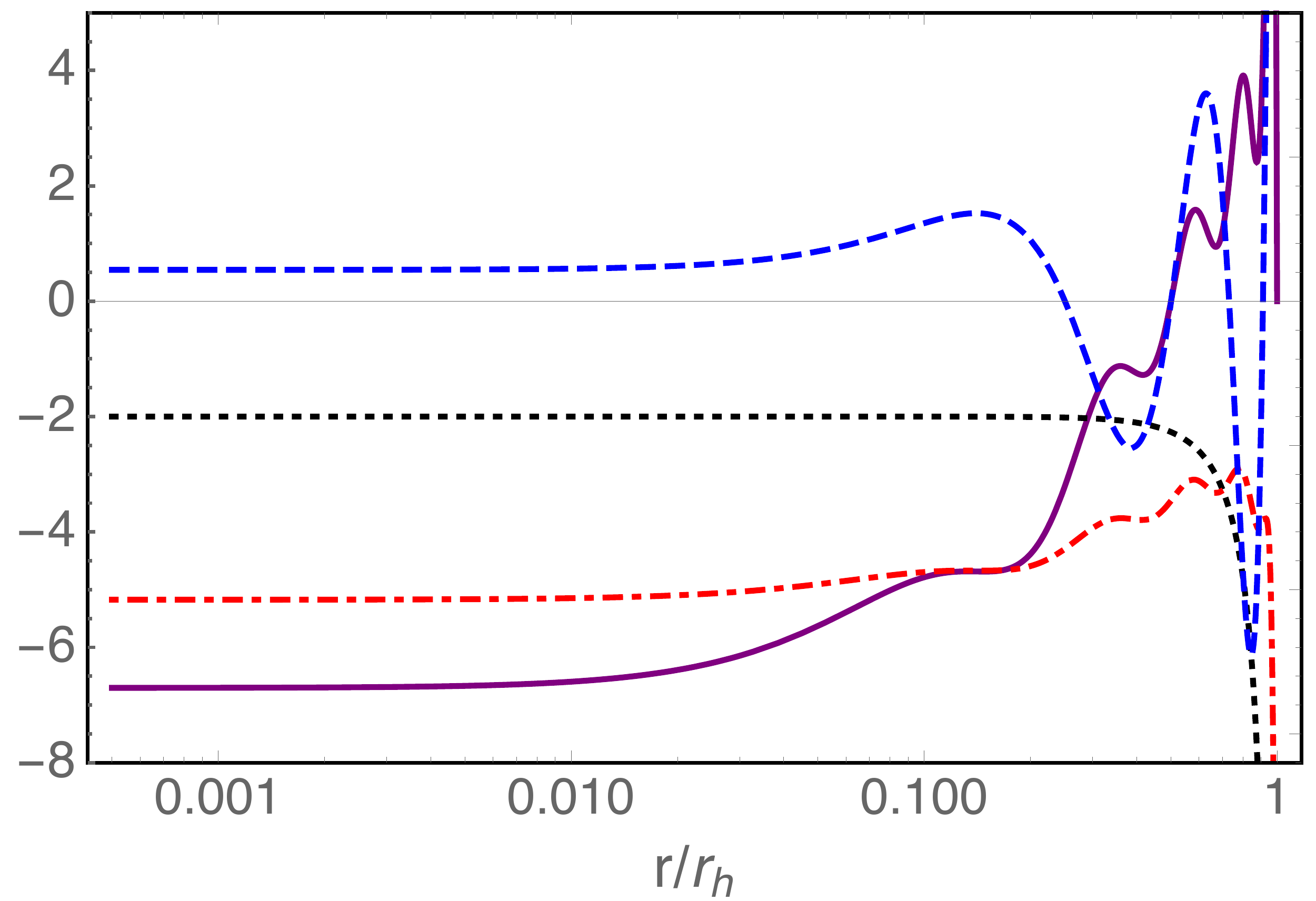}~~~~~~
\includegraphics[
width=0.32\textwidth]{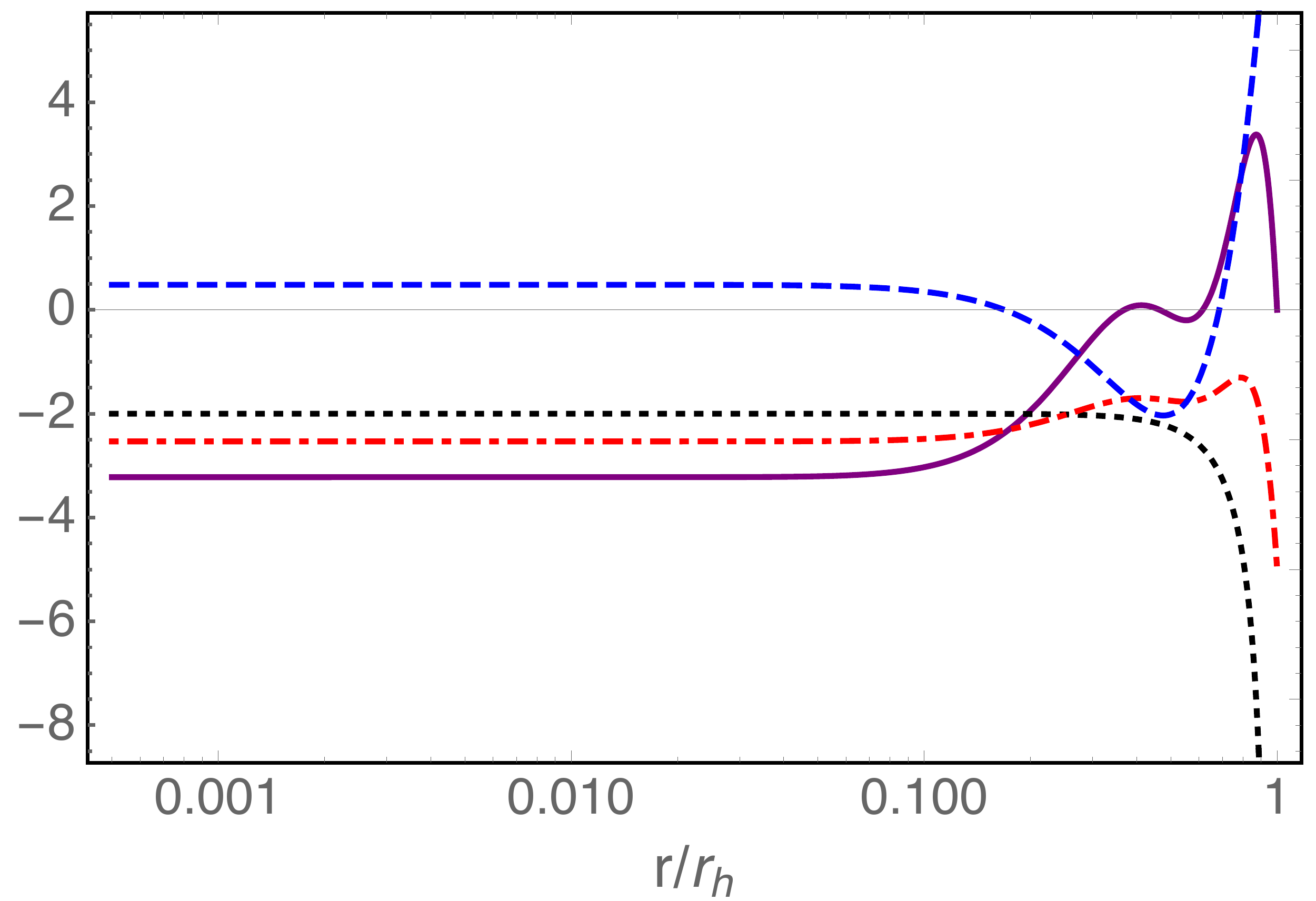}~~~~~~
\includegraphics[
width=0.32\textwidth]{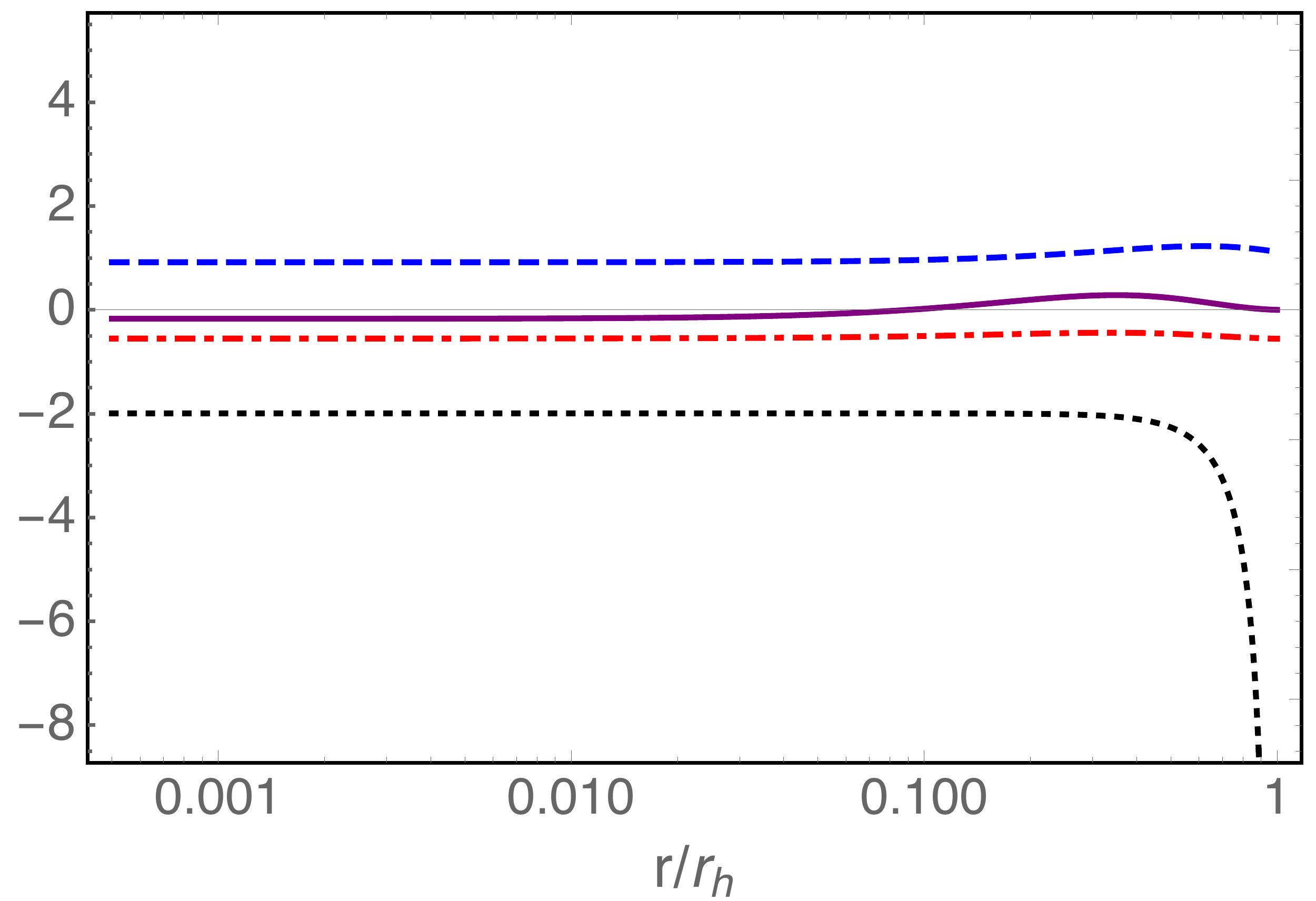}
\end{center}
\vspace{-0.5cm}
\caption{\small Three typical examples  of the configuration in the neutral helical black hole interior with fixing $\alpha_b=0.1$ at different temperatures $T/k=0.05$ ({\em left}), $0.1$ ({\em middle}) and $0.3$ ({\em right}). The functions we plot are $rf'/(\alpha_h^2 f)
$ (purple), $
r \alpha'/\alpha_h
$ (blue), $
(rh'/h-1)/(4 \alpha_h^2)
$ (red), 
$
rg'/g
$ (black). } 
\label{fig:con-neu}
\end{figure}

With the above setups, we can numerically solve the system for given $T/k$ and $\alpha_b$ from the AdS boundary to the black hole singularity. In Fig. \ref{fig:con-neu}, we show three typical configurations of the full dynamical evolution for the metric fields inside the horizon, where we have fixed $\alpha_b=0.1$ and increase the temperature $T/k$ from $0.05$ in the left plot to $0.1$ in the middle plot, to $0.3$ in the right plot. Note that $r/r_h=1$ is the horizon and $r/r_h\to 0$ is the singularity. We have numerically solved the system from the horizon to $r/r_h\sim 10^{-90}$ although we only show part of them in the figure since the configurations from $10^{-4}$ to $10^{-90}$ are trivial constants. We have used the radial conserved quantity \eqref{eq:neucc} to check the accuracy of our numerics. We find that at all the different temperatures, evolving from the horizon to the singularity, the fields evolve after an oscillate regime or smoothly to the stable Kasner epoch. When the temperature is low enough,  the curves related to fields $\alpha, f, h$ has an oscillation regime close to the horizon.\footnote{This reminds us the Josephson oscillation for the scalar field in the interior of holographic superconductor \cite{Hartnoll:2020fhc}. 
For the neutral helical black holes there is no background phase winding phenomenon and we have not been aware of any analogous Josephson effect.  Therefore we call this just oscillation.} At higher temperature, the oscillation behavior disappears. Note that in previous studies, no oscillation behavior was found for neutral black holes with scalar deformations \cite{Frenkel:2020ysx}. In addition,  the Kasner exponents are stable and we have not found any Kasner inversion or Kasner transition for these neutral helical black holes. 

\begin{figure}[h!]
\begin{center}
\includegraphics[
width=0.43\textwidth]{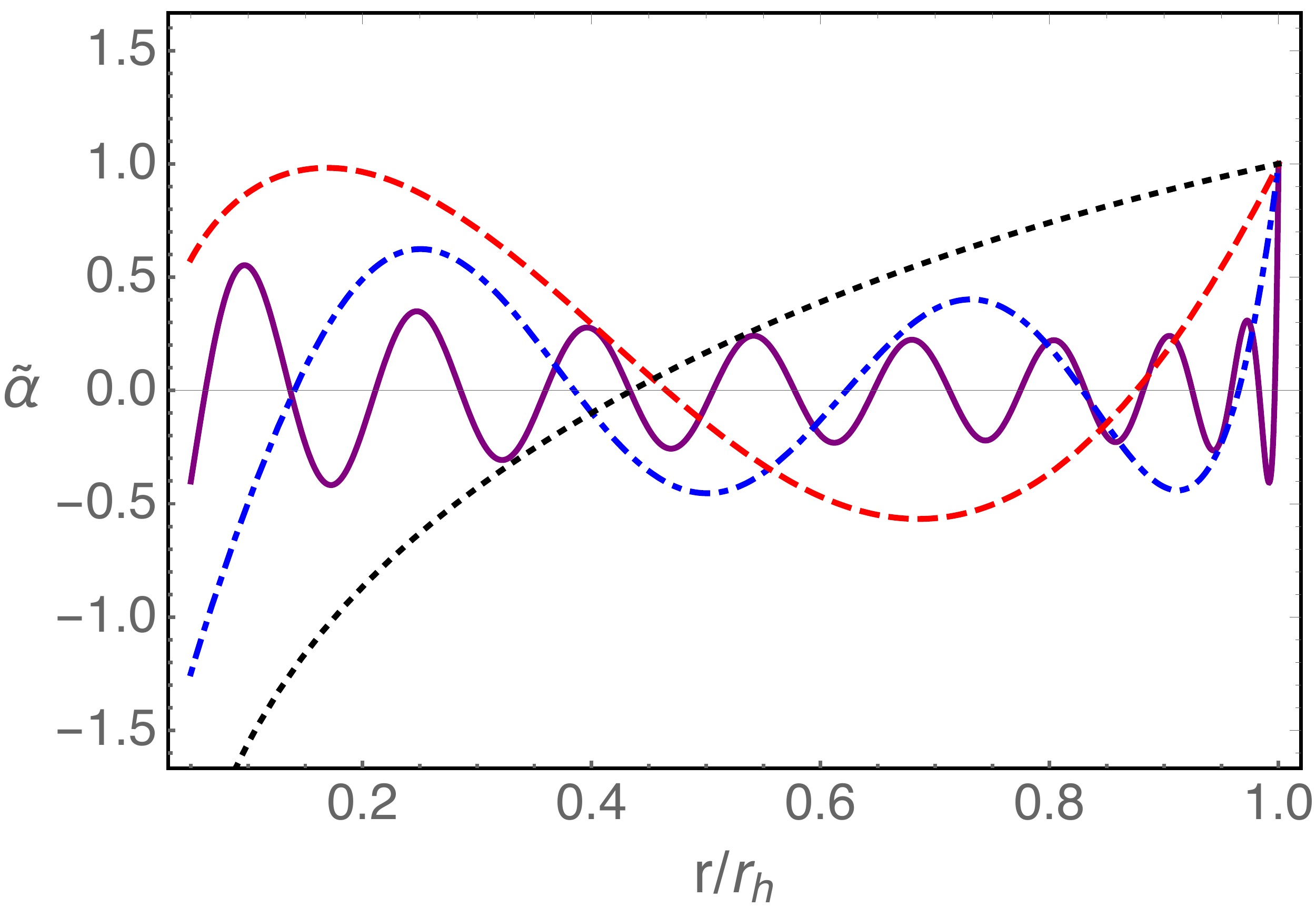}\,~~~~~~
\includegraphics[
width=0.43\textwidth]{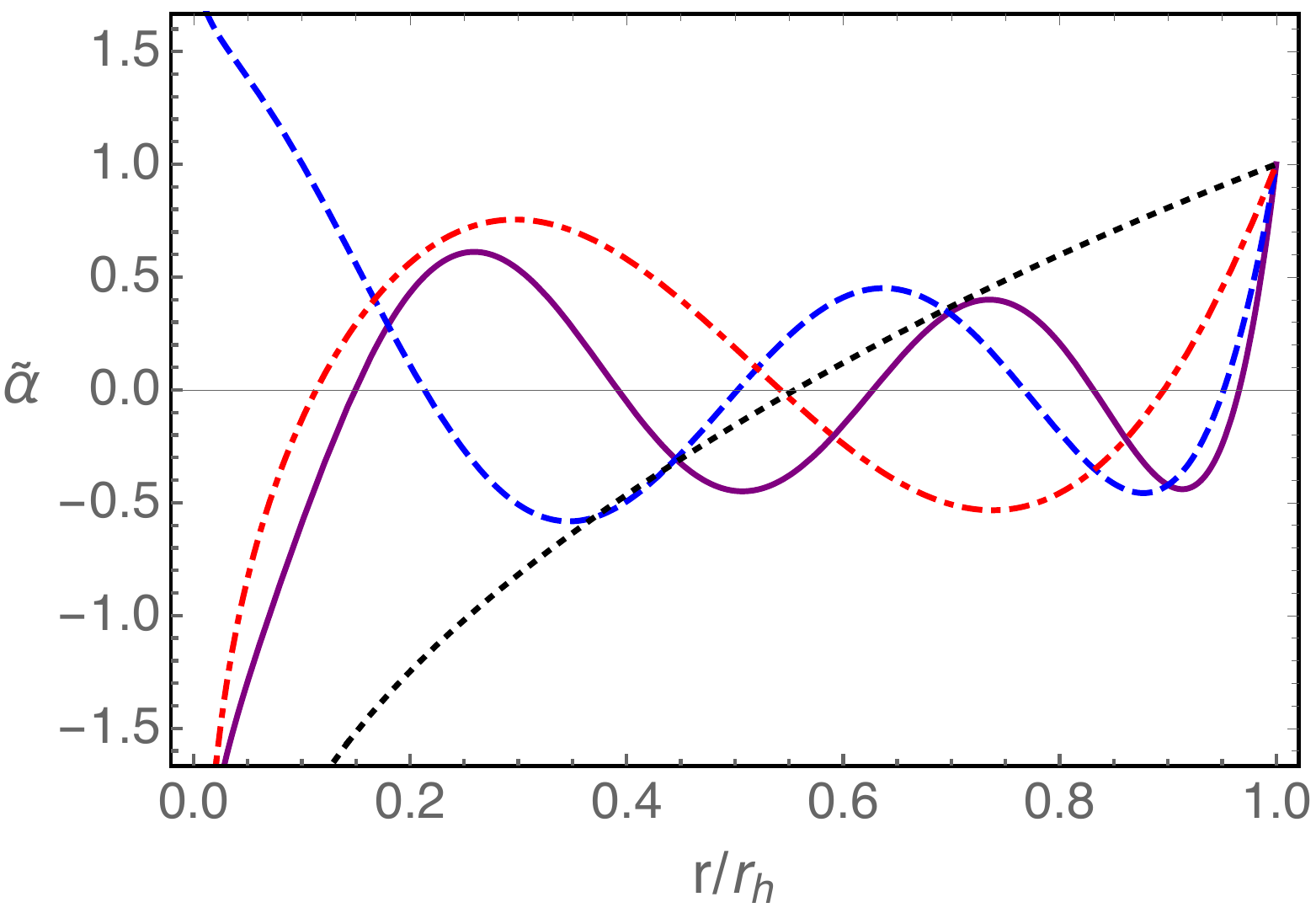}
\end{center}
\vspace{-0.7cm}
\caption{\small The plots of $\tilde \alpha=\alpha/\alpha_h$ along radial direction in the oscillation region for fixing $\alpha_b=0.1$ ({\em left}) while 
$T/k =0.015$ (purple), $0.05$ (blue), $0.1$ (red),  $0.3$ (black), as well as for fixing $T/k=0.05$ ({\em right}) while 
$\alpha_b =0.1$ (purple), $0.3$ (blue), $0.5$ (red),  $1$(black). Here $\alpha_h$ is the horizon value of $\alpha$. 
} 
\label{fig:neualpha}
\end{figure}

In Fig. \ref{fig:neualpha}, we show the oscillation regime of the field $\alpha$ (which has been rescaled according to $\tilde \alpha=\alpha/\alpha_h$) as a function of $r/r_h$ for fixing $\alpha_b$ (left) or $T/k$ (right) respectively. The left plot in Fig. \ref{fig:neualpha} shows that for fixing $\alpha_b$, 
the field $\alpha$ oscillates more dramatic at lower temperature, and the right plot shows that for fixing temperature the smaller $\alpha_b$ is, the more times $\alpha$ oscillates.  We also find similar  oscillation behavior at low temperature for the  fields $rf'/f$ and $rh'/h$ which can be seen from Fig. \ref{fig:con-neu}.\footnote{We have checked that there is no oscillation behavior for the metric fields outside the horizon for the parameters.  Additionally, the frequency of oscillations for $rf'/f$ and $rh'/h$ are roughly two times of $r\alpha'$.} It is interesting to note that at low temperature inside the horizon, the spatial components of the metric $e^{\pm2\alpha}$ in \eqref{eq:ans1} 
show oscillation behavior before reaching Kasner regime.
Our results also suggest that the oscillation behavior near the horizon is not necessarily to be in contact with the collapse of the Einstein-Rosen bridge. It remains to be seen that if there exists an analytical way to calculate the fields in the oscillation regime and the holographic signature of the oscillation behavior in the dual field theory. 

\subsection{Kasner exponents}
After the oscillation regime, the fields evolve to the Kasner epoch which can be studied analytically. Near the singularity, i.e. $r\to 0$, we assume
\begin{align}
\label{eq:scalingsol}
\begin{split}
    g= -\frac{g_0}{r^{n_g}}+\dots\,, ~~~f= f_0r^{n_f}+\dots\,, ~~~h= h_0 r^{n_h}+\dots\,, ~~~\alpha= n_{\alpha} \log r + \alpha_0+\dots\,
    \end{split}
\end{align}
where $g_0, n_g, f_0, n_f, h_0, n_h, n_\alpha, \alpha_0$ are all constants, and the ``$\dots$" are subleading in $r$ compared to the first terms. 

When $r\to 0$, 
the equations of motion \eqref{eq:eomnetural} can be simplified as
\begin{align}
\label{eq:rto0n}
\begin{split}
    f'+\frac{f(hh'-rh'^2-2rh^2\alpha'^2)}{h(2h+rh')}&=0\,,\\
    g'+\frac{2g(rhh'+r^2h'^2+h^2(1+r^2\alpha'^2))}{rh(2h+rh')}&=0\,,\\
    h''-\frac{h'^2}{h}+\frac{h'}{r}&=0\,,\\
    \alpha''+\frac{\alpha'}{r}  &=0\,,
\end{split}
\end{align}
from which we obtain
\begin{align}
\label{eq:nfngn}
    n_f=\frac{n_h(n_h-1)+2n_{\alpha}^2}{n_h+2}\,,~~~~~ n_g=\frac{2(n_h^2+n_h+n_{\alpha}^2+1)}{n_h+2}\,.
\end{align}
Note that in above equations \eqref{eq:rto0n}, we have assumed that the terms ignored should be subleading. More explicitely, we have assumed
\begin{align}
\label{eq:assneu}
    n_g>-2\,, ~~~n_g-2n_h+2>0\,,  ~~~n_g-2n_h+2\pm 4n_{\alpha}>0\,.
\end{align}
Combing \eqref{eq:assneu} and \eqref{eq:nfngn}, we have
\be 
\label{eq:con1}
n_h>-2, ~~~~n_g\geq 4\sqrt{3}-6\,,~~~~n_\alpha^2+3\pm 2n_\alpha(n_h+2)>0\,.\ee

Evaluate the conserved charge \eqref{eq:neucc} at the  singularity and the horizon, we obtain 
\begin{align}
\begin{split}
   & \frac{4f_0 g_0 h_0 (3n_h - n_{\alpha}^2 +3)}{n_h+2} = 8f_h h_h r_h^3=2 f_b h_b Ts \,. 
    \end{split}
\end{align}
Given the fact that the term on the right hand side is positive and $f_0, g_0, h_0$ are all positive, we have constraint 
\be
\label{eq:con2}
3n_h+3>n_\alpha^2\,, \ee 
from which we further have $n_h>-1$. 

Although we could not further constrain these parameters, it is necessarily that all the inequalities should be satisfied. It is easy to see there always exist a range of values $n_\alpha, n_h$ for which the constraints in \eqref{eq:con1} and \eqref{eq:con2} are all  satisfied. Numerically we have checked that the values we obtained satisfy the constraints. 
This indicates the Kasner regime is stable and we have not found any Kasner inversion or transition in this model. 
The stability of Kasner geometry is quite similar to the neutral black holes with scalar hair in \cite{Frenkel:2020ysx,Wang:2020nkd, Mansoori:2021wxf,Liu:2021hap}.

Starting from \eqref{eq:scalingsol} and performing the coordinate transformation 
$\tau=\frac{2}{\sqrt{g_0}(n_g+2)}r^{(n_g+2)/2}$, the metric near the singularity becomes the Kasner form \cite{Kasner:1921zz}
\be
\label{eq:kasner-neu}
ds^2= -d\tau^2+ c_t \tau^{2p_t}dt^2 + c_1 \tau^{2p_1}\omega_1^2 + c_2 \tau^{2p_2}\omega_2^2 + c_3 \tau^{2p_3}\omega_3^2 
\ee
with the Kasner exponents 
\begin{align}
    p_t=\frac{2n_f-n_g}{n_g+2}\,,~~ p_1=\frac{2n_h}{n_g+2}\,,~~p_2=\frac{2(n_{\alpha}+1)}{n_g+2}\,,~~p_3=\frac{2(1-n_{\alpha})}{n_g+2}\,.
\end{align}
Substituting \eqref{eq:nfngn} into the above equations, we find  
\begin{align}
\begin{split}
    p_t=\frac{n_{\alpha}^2-2n_h-1}{n_{\alpha}^2+n_h(n_h+2)+3}&\,,~~~ p_1=\frac{n_h(n_h+2)}{n_{\alpha}^2+n_h(n_h+2)+3}\,,\\
    p_2=\frac{(n_h+2)(n_{\alpha}+1)}{n_{\alpha}^2+n_h(n_h+2)+3}&\,,~~~p_3=\frac{(n_h+2)(1-n_{\alpha})}{n_{\alpha}^2+n_h(n_h+2)+3}\,.
    \end{split}
\end{align}

From the above equations, one can easily check
\begin{align}
\label{eq:kasnerrelation-neu}
    p_t+p_1+p_2+p_3=p_t^2+p_1^2+p_2^2+p_3^2=1\,.
\end{align}
Note that different from the neutral black holes in four dimensions  \cite{Frenkel:2020ysx}, here we have two independent Kasner exponents for the singularity inside the five dimensional helical black holes. An immediate question is how to describe these two independent Kasner exponents from the dual field theory. It seems that we need two different observables at the same time. We leave this interesting question for future study. 

\begin{figure}[h!]
\begin{center}
\includegraphics[
width=0.25\textwidth]{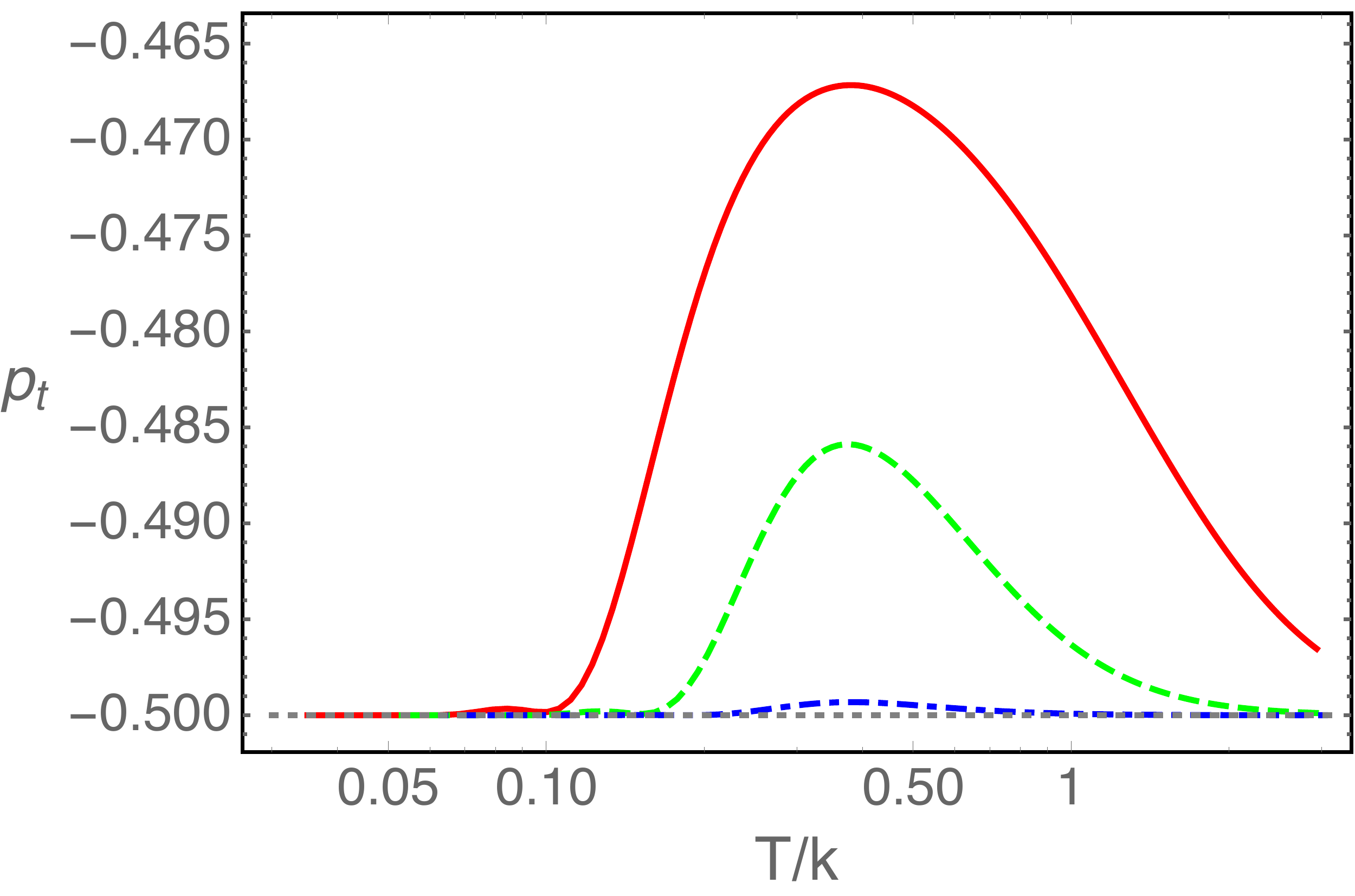}\,~~~
\includegraphics[
width=0.245\textwidth]{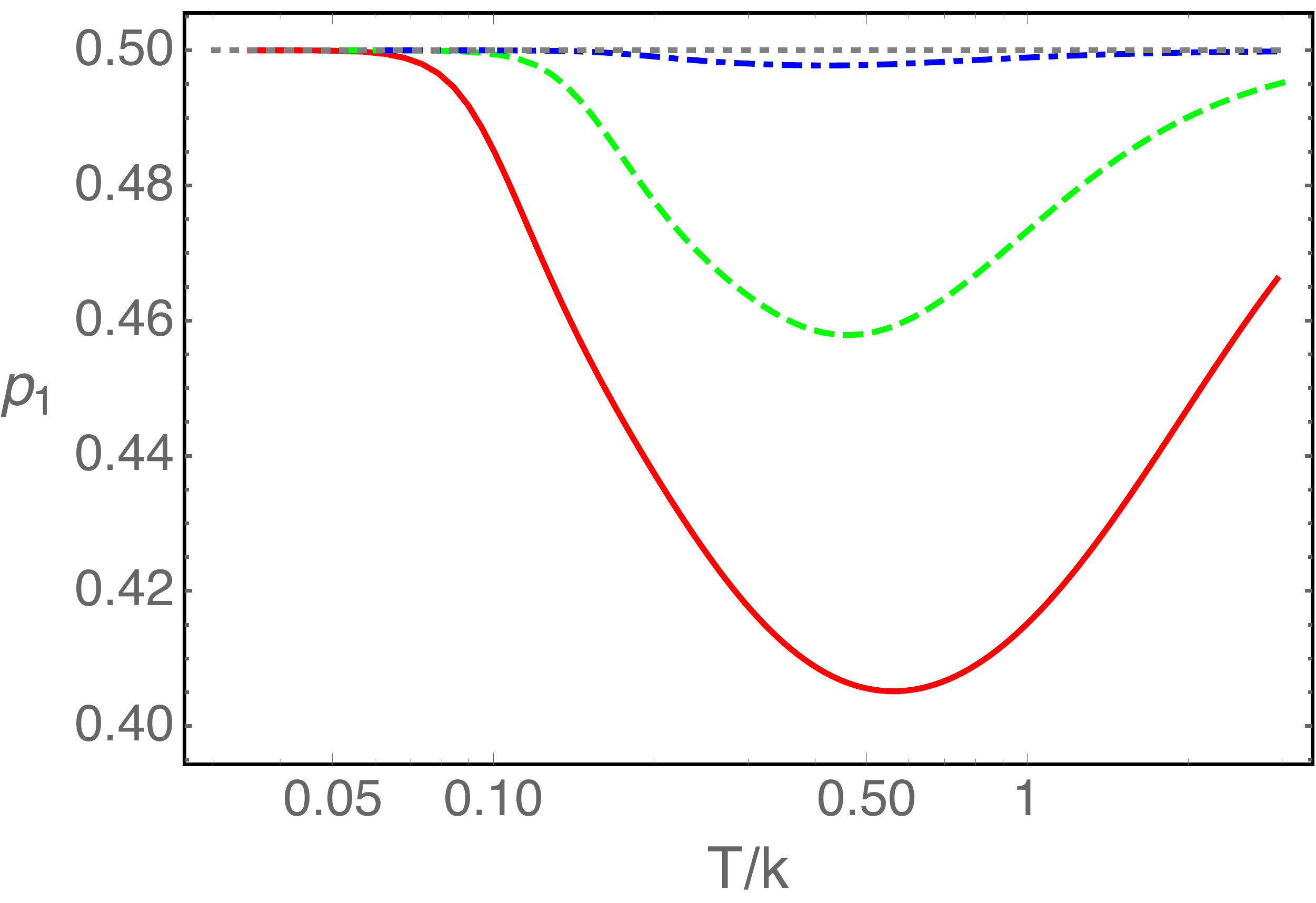}\,~~~
\includegraphics[
width=0.245\textwidth]{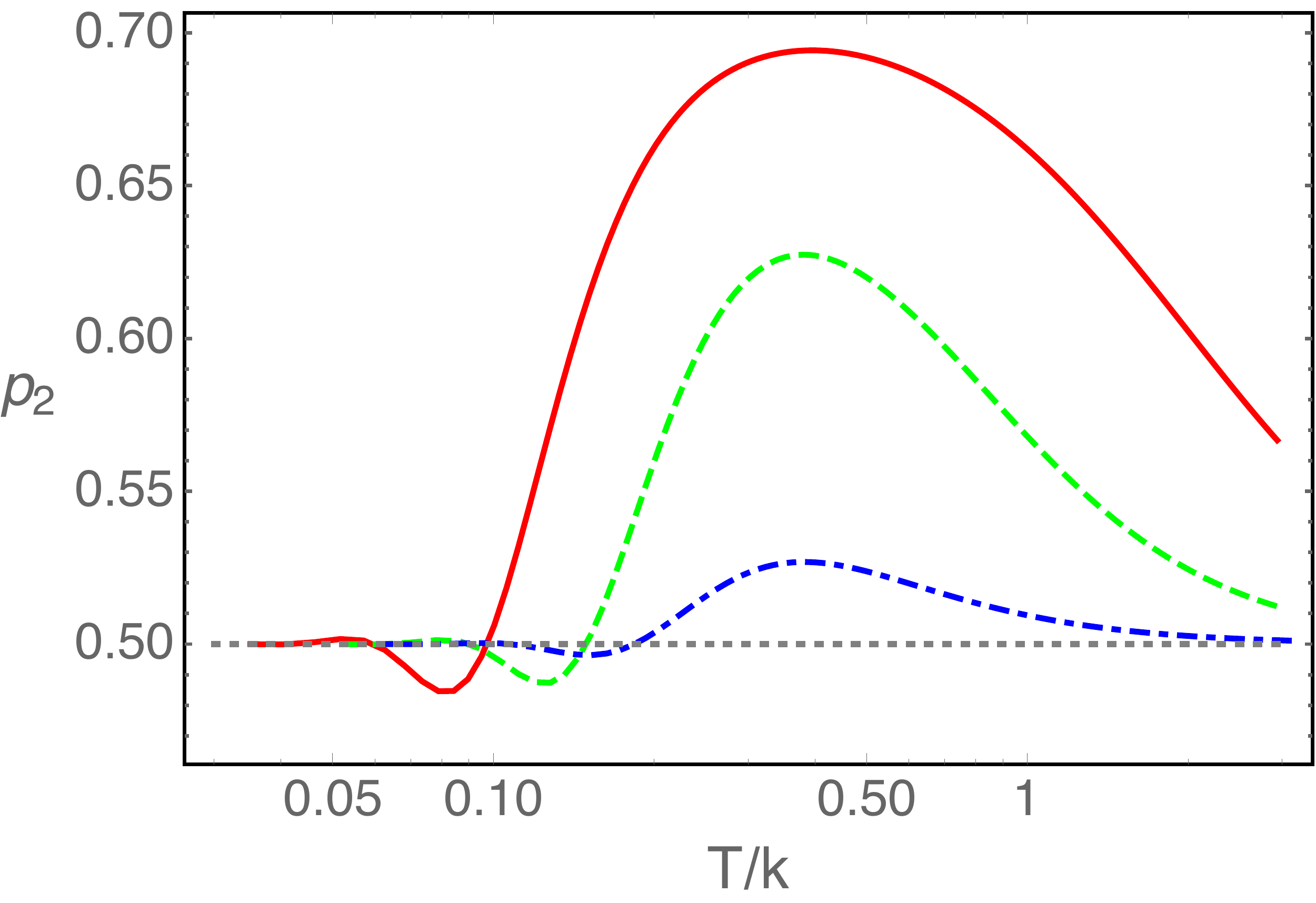}\,~~~
\includegraphics[
width=0.245\textwidth]{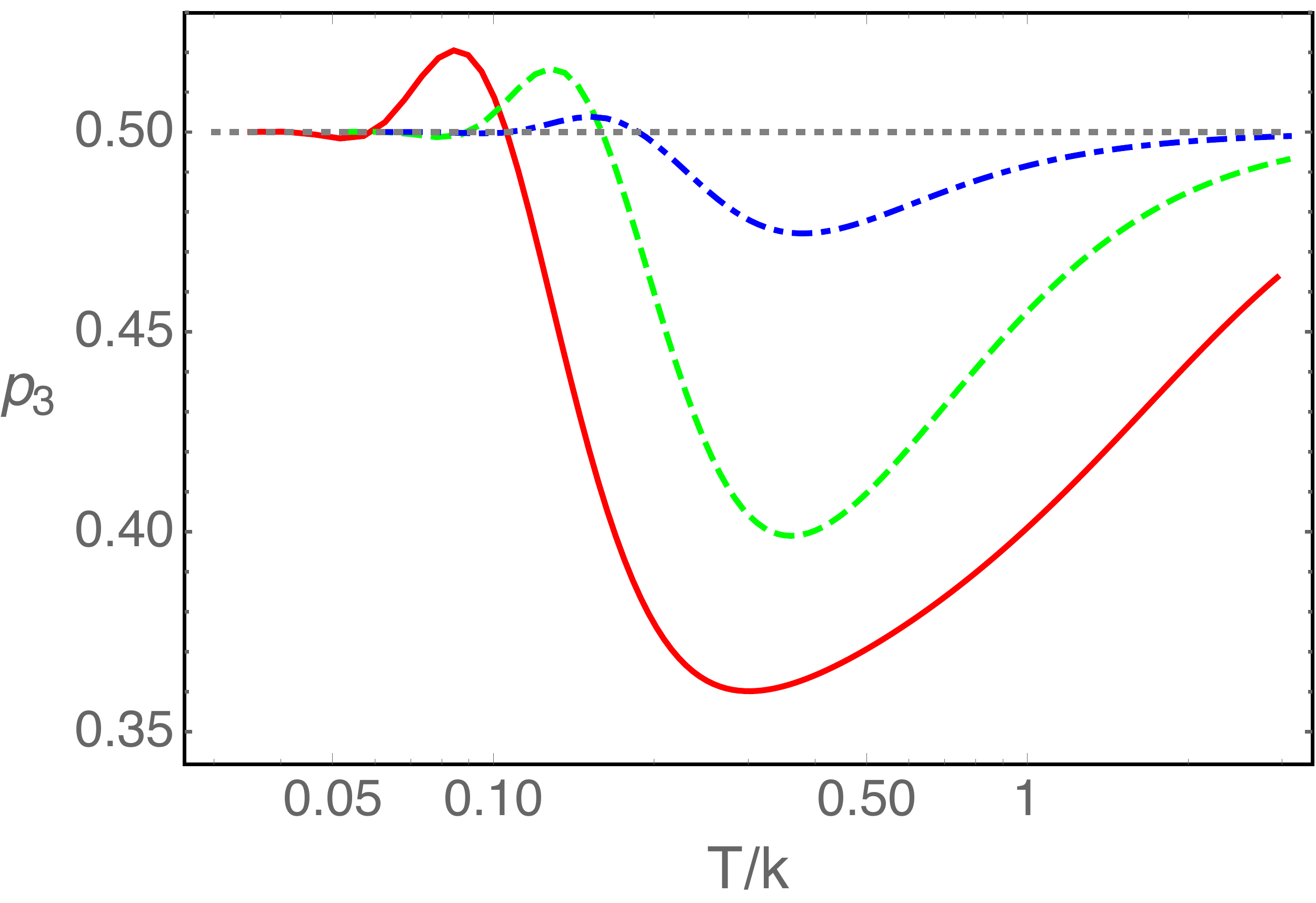}
\end{center}
\vspace{-0.5cm}
\caption{\small The Kasner exponents as a function of $T/k$ 
for 
$\alpha_b =1$ (red), $0.5$ (green), $0.1$ (blue). The gray dotted line is for Schwarzschild black hole with $p_t=-0.5=-p_1=-p_2=-p_3$.} 
\label{fig:neuKasT}
\end{figure}

The Kasner exponents can be obtained numerically.  
In Fig. \ref{fig:neuKasT}, we have shown the four  Kasner exponents in \eqref{eq:kasner-neu} as functions of $T/k$ at different $\alpha_b$. 

\begin{figure}[h!]
\begin{center}
\includegraphics[
width=0.25\textwidth]{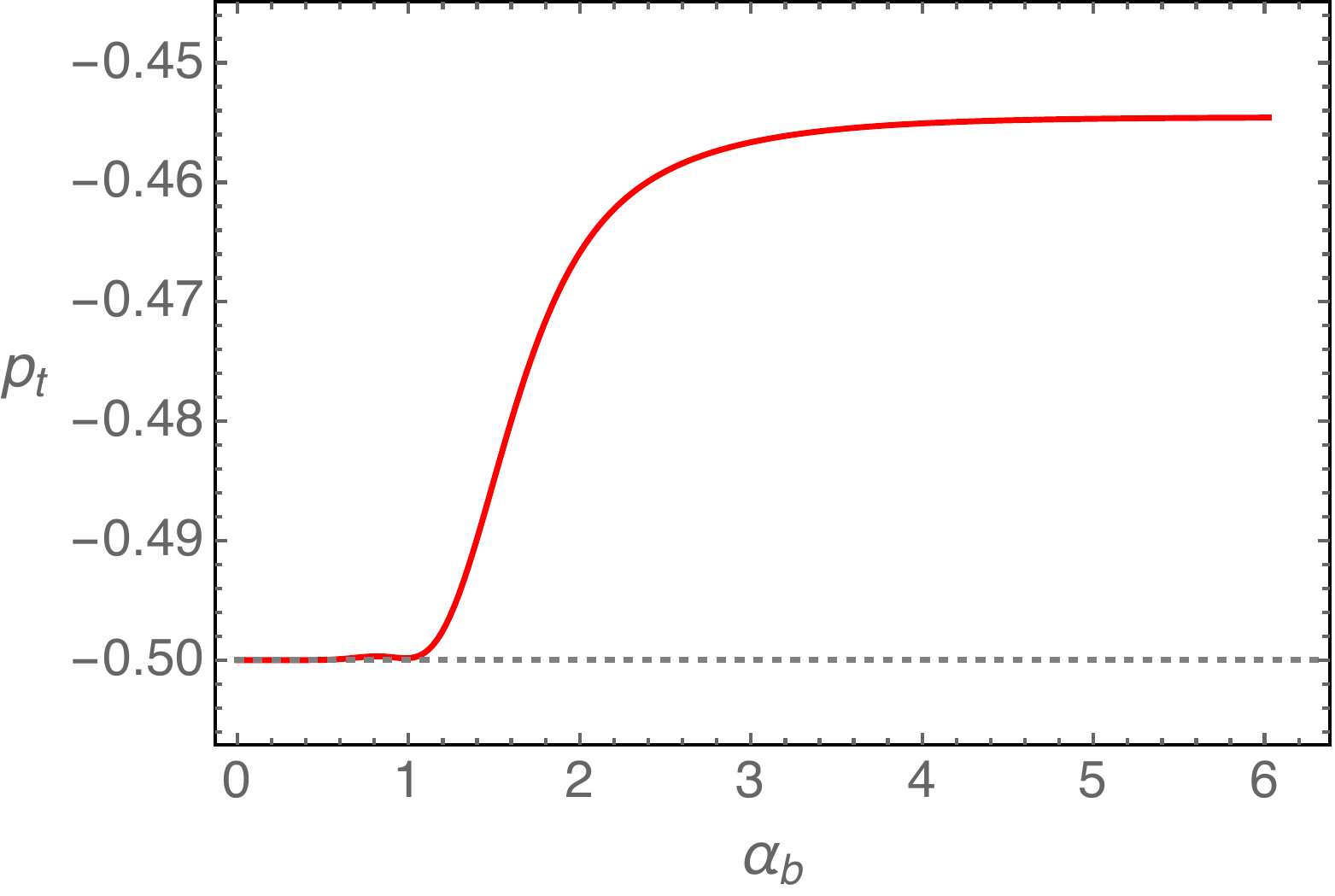}\,~~~
\includegraphics[
width=0.244\textwidth]{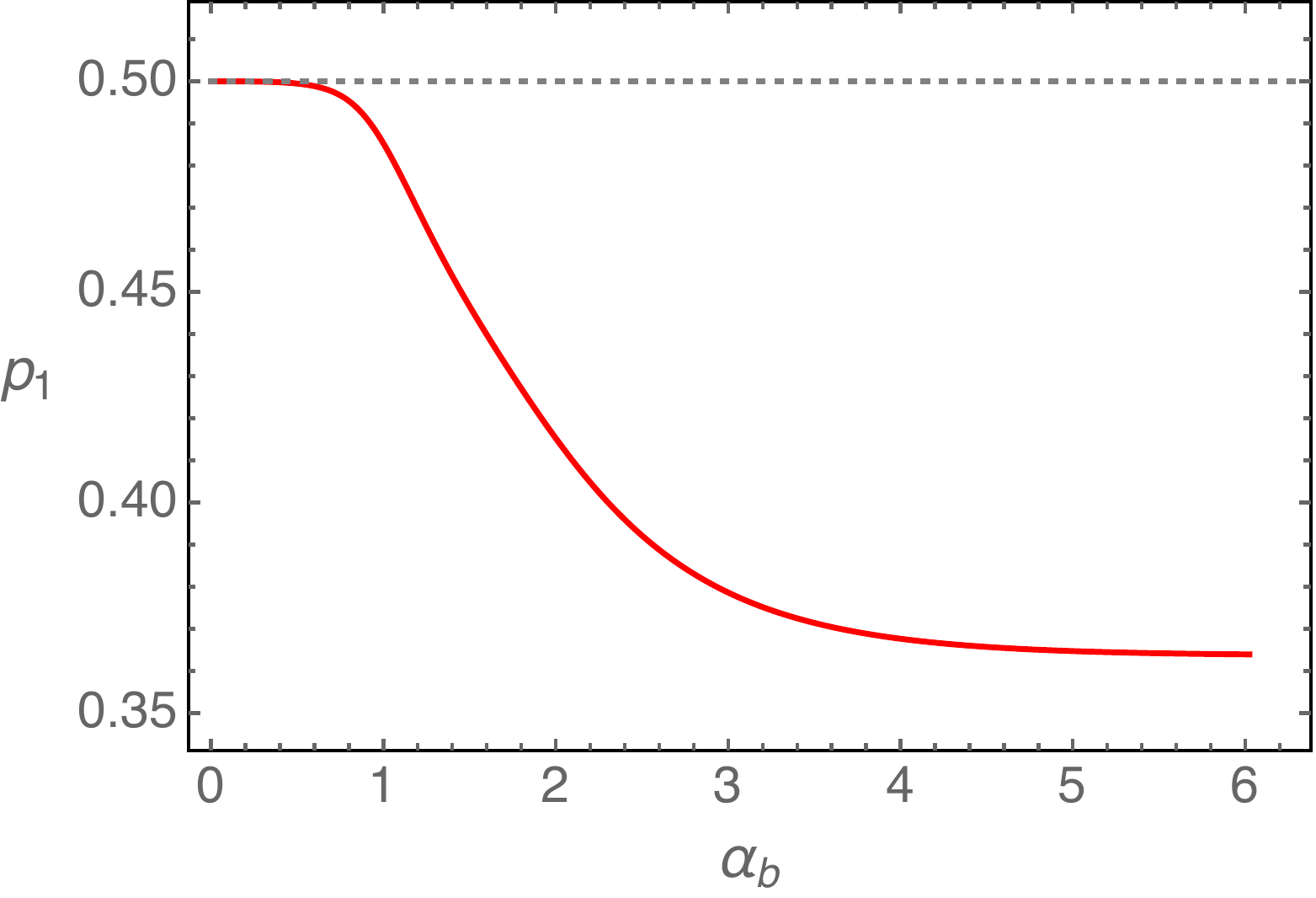}\,~~~
\includegraphics[
width=0.244\textwidth]{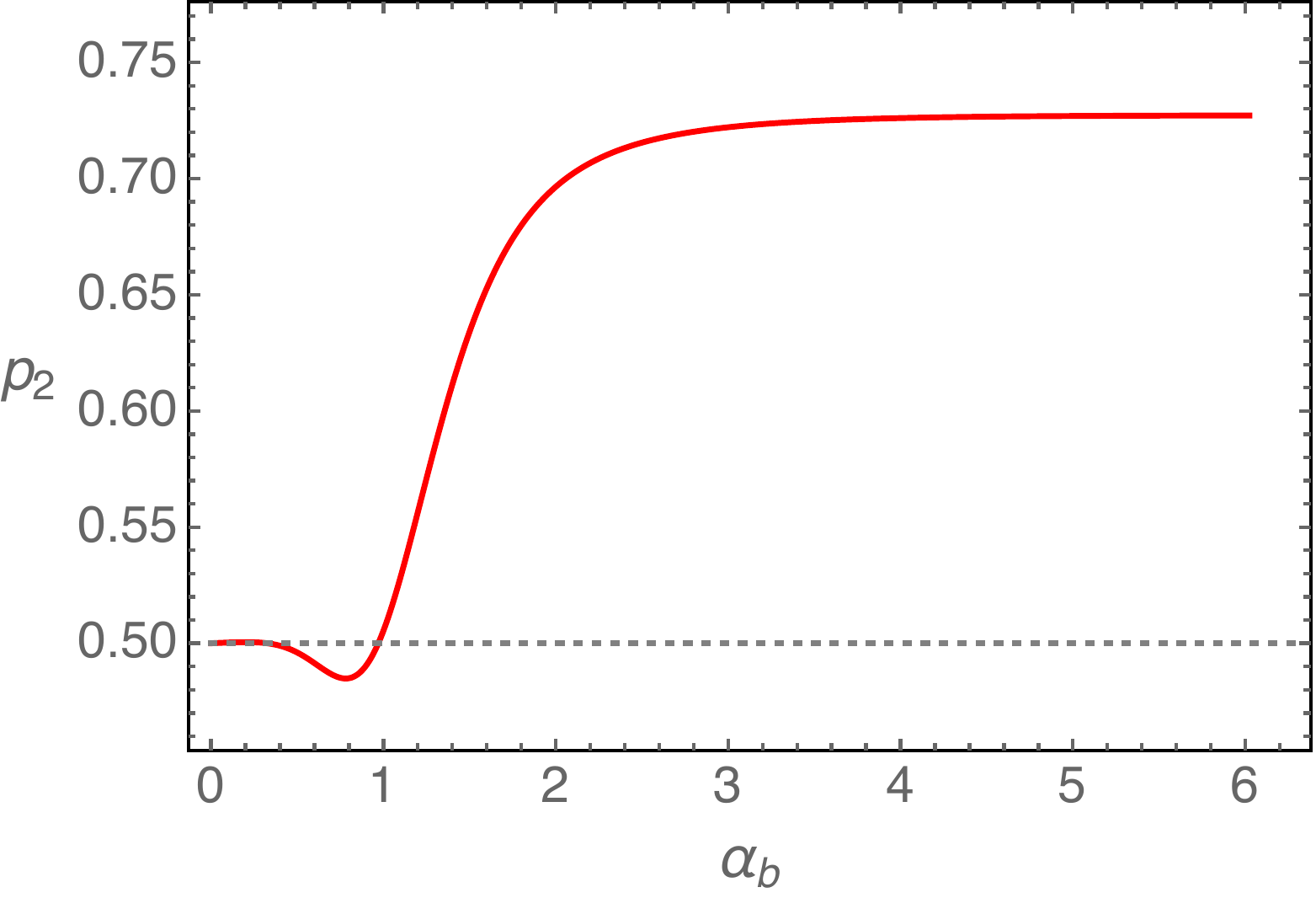}\,~~~
\includegraphics[
width=0.244\textwidth]{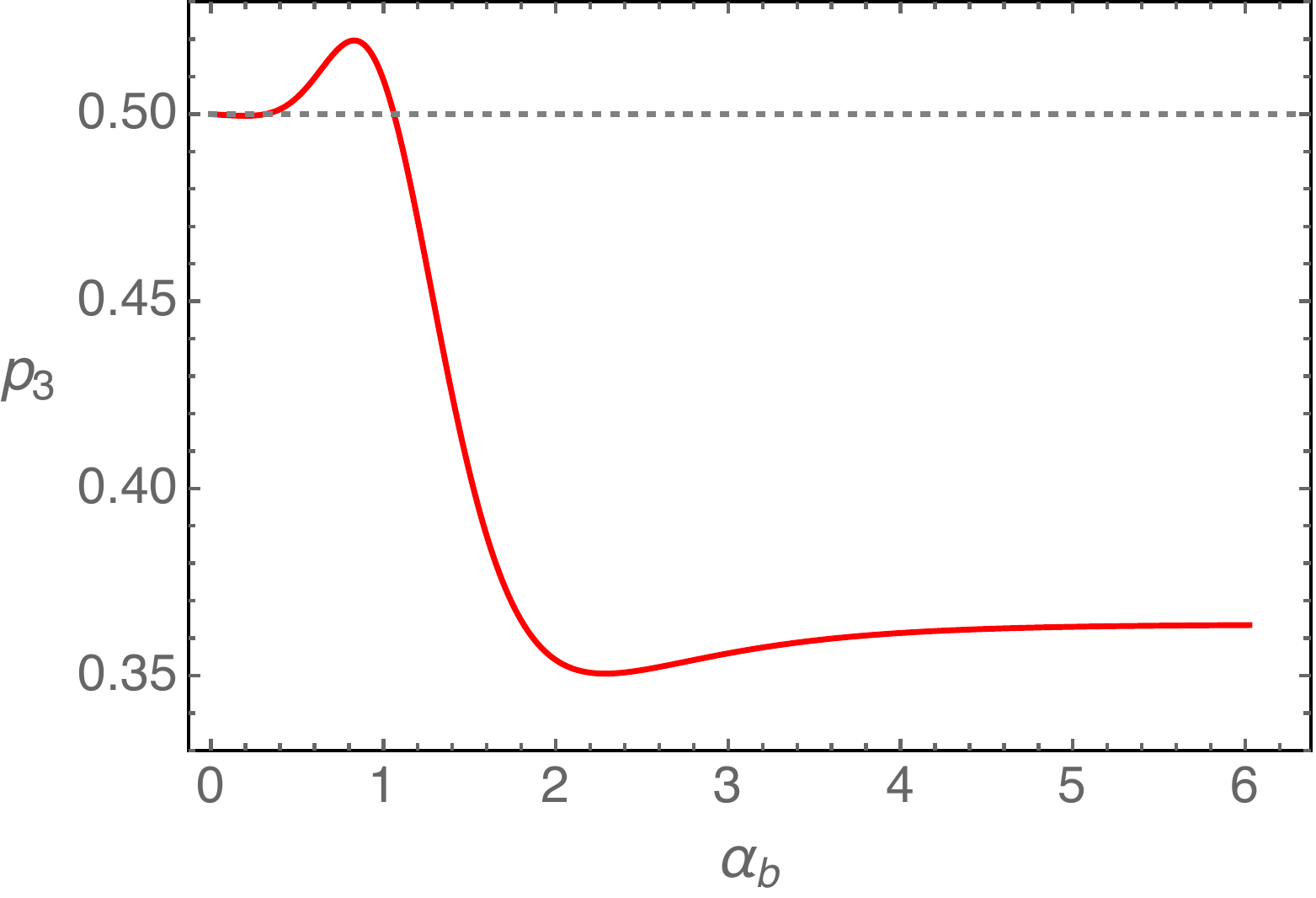}
\end{center}
\vspace{-0.7cm}
\caption{\small The Kasner exponents as functions of $\alpha_b$ at $T/k=0.1$. The gray dotted line is corresponding value of Schwarzschild black hole.} 
\label{fig:neuKasalpha}
\end{figure}

In Fig. \ref{fig:neuKasalpha}, we show the dependence of Kasner exponents as functions of $\alpha_b$ for fixed temperature $T/k=0.1$. We have numerically check that the Kasner relations \eqref{eq:kasnerrelation-neu} are satisfied for all the configurations we have considered. Furthermore, numerically for neutral helical black holes we only find the case with $p_t<0$ while the other Kanser exponents are positive.

\section{Charged helical black holes}\label{sec3}

Comparing to the neutral Schwarzschild black holes with scalar deformation, the interior dynamics of charged  Reinssner-Nordstr\"{o}m black holes with scalar deformation is much more fruitful \cite{Frenkel:2020ysx,Hartnoll:2020rwq,Hartnoll:2020fhc}. Similarly, one might expect that 
the charged helical black holes contain quite  interesting interior dynamics. 

In this section we consider the interior structure of five dimensional helical black holes at finite density in Einstein-Maxwell gravity. The action is 
\begin{align}
\begin{split}
\label{eq:acc}
  S=\int d^5x \sqrt{-g} \bigg(R+12-\frac{1}{4}F^2\bigg)\,.
  \end{split}
\end{align}
The equations of motion are
\begin{align}
\label{eq:eom-ch}
 R_{ab}+4g_{ab} &=\frac{1}{2}F_{ac}F_b^{\ c}-\frac{1}{12}g_{ab}F_{cd}F^{cd}\,,\\   
    \nabla_{b}F^{ba}&=0\,.
    \label{eom:maxwell}
\end{align}

The ansatz for the background fields are
\begin{align}
\begin{split}
    ds^2&=-gf^2 dt^2+\frac{dr^2}{g}+h^2 \omega_1^2+r^2 e^{2\alpha}\omega_2^2+r^2 e^{-2\alpha}\omega_3^2\,,\\
  A&=\psi dt\,,\label{ansatz:em}
  \end{split}
\end{align}
where $f,g,h,\alpha,\psi$ are functions of $r$. The metric part is the same as \eqref{eq:ans1} for neutral helical black holes and here we have an additional gauge potential. $\omega_i$ are one-forms defined in \eqref{eq:oneform} satisfying Bianchi VII$_0$ algebra. The dual field theory is located at $r\to\infty$. The horizon and the singularity are located at $r=r_h$ and $r\to 0$ respectively. 

Substituting the ansatz \eqref{ansatz:em} into EOM, we obtain
\begin{align}
\begin{split}
     \frac{f'}{f}+ \frac{(4r^2+g)hh'+r(2k^2\sinh^2{2\alpha}-gh'^2)-2rh^2(g\alpha'^2+2)}{gh(2h+rh')} + \frac{r\psi'^2(h-rh')}{6f^2g(2h+rh')}&=0\,,\\
     \frac{g'}{g} + \frac{2\left(r(g-4r^2)hh'-r^2(k^2\sinh^2{2\alpha}-gh'^2) + h^2(g+r^2g\alpha'^2 - 2r^2)\right)}{rgh(2h+rh')} + \frac{r\psi'^2(h+2rh')}{6f^2g(2h+rh')}&=0\,,\\
     h''+\frac{h'}{r} -\frac{h'^2}{h} +\frac{2k^2\sinh^2{2\alpha}}{gh} + \frac{1}{g} \left(\frac{\psi'^2}{6f^2}-4\right) (h-rh')&=0\,,\\
    \alpha''+\frac{\alpha'}{r} +\frac{4r\alpha'}{g} - \frac{r\alpha'\psi'^2}{6f^2g} -\frac{k^2 \sinh{4\alpha}}{gh^2}&=0\,,\\
    \psi''-\psi'\left(\frac{f'}{f}-\frac{h'}{h}-\frac{2}{r}\right)&=0\,.\label{eq:ceom}
    \end{split}
\end{align}
When $\psi=0$, the above equations reduce to the neutral case with equations \eqref{eq:eomnetural}. 
Furthermore, the last equation can be simplified to 
$
    \left(\frac{r^2h\psi'}{f}\right)'=0.
$ 
This implies
\begin{align}
\label{eq:psi}
    \psi'=\frac{2 f}{r^2 h}\psi_0\,, 
\end{align}
where $\psi_0$ is the integration constant which could be interpreted as the charge density of the dual field theory. 
Plugging  \eqref{eq:psi} into equations of motion \eqref{eq:ceom}, we obtain four equations for functions $f,g,h,\alpha$.

Three different sets of scaling symmetries of the ansatz \eqref{ansatz:em} are
\begin{align}
    &r\rightarrow \lambda r\,,~~~ \{t,x_2,x_3\}\rightarrow \lambda^{-1}\{t,x_2,x_3\}\,,~~~g\rightarrow \lambda^2 g\,,~~~\psi\rightarrow \lambda\psi\,;\label{eq:csym1}\\
    &t\rightarrow \lambda t\,,~~~ f\rightarrow \lambda^{-1} f\,,~~~ \psi\rightarrow \lambda^{-1} \psi\,;\label{eq:csym2}\\
    &x_1 \rightarrow \lambda^{-1} x_1\,,~~~ h\rightarrow \lambda h\,,~~~ k\rightarrow \lambda k\,.\label{eq:csym3}
\end{align}
These scaling symmetries are important to numerically solve the system. A useful radial conserved quantity associated to the above symmetries is 
\begin{align}
\label{eq:ncc}
    Q=\frac{2r^4h}{f}\left(\left(\frac{f^2g}{r^2}\right)'-\frac{\psi\psi'}{r^2} \right)\,.
\end{align}
This conserved quantity can be used to check our numerical calculations. 

The near horizon and near boundary expansions of the fields can be found in appendix \ref{app:b}. It turns out the bulk geometry is completely determined by three dimensionless parameters $T/\mu$, $k/\mu$ and $\alpha_b$, where $T$ and $\mu$ are the temperature and the chemical potential of the charged helical black hole, the wave-number $k$ characterizes the pitch of the helical structure which is equal to $2\pi/k$ and $\alpha_b$ is the strength of the helical deformation. 

The thermodynamic quantities of the system are 
\begin{align}
    T=\frac{r_h(24 f_h^2 -\psi_1^2) }{24\pi f_b f_h} = \frac{f_h(6h_h^2 r_h^4 -\psi_0^2)}{6f_b h_h^2 r_h^3}\,,~~~~ s=\frac{4\pi r_h^2 h_h}{h_b}\,.
\end{align}

One can check that the equations \eqref{eq:ceom} have solutions as Reissner–Nordstr\"{o}m black hole
when $\alpha=0$, 
\begin{align}
\begin{split}
    f&=1\,,~~~~h=r\,,\\
    g&=r^2-\frac{r_h^2(3r_h^2+\mu^2)}{3r^2}+\frac{\mu^2 r_h^4}{3r^4}\,,\\
    \psi&=\mu\left(1-\frac{r_h^2}{r^2}\right)\,.
\end{split}
\end{align}
When we have a non-trivial source $\alpha_b$, the field $\alpha$ has non-trivial profile. 

In the following we will numerically solve the system and study the interior structure of the charged helical black holes.

\subsection{Non-existence of inner horizon}
We first give a proof that there could not be any inner horizon for the charged helical black holes \eqref{ansatz:em}. The proof of no-inner horizon theorem for charged black hole in presence of charged scalar field or vector field can be found in \cite{Hartnoll:2020fhc, Cai:2020wrp, Cai:2021obq}.  Although our setup is different since the deformation is driven by the metric field $\alpha$, we could use similar strategy to prove it. 

The proof of no inner horizon can be obtained from the equation of motion for $\alpha$. First, we note that a compact equation of $\alpha$ can be obtained by a linear combination of the first, second and fourth equations in \eqref{eq:ceom},  
\begin{align}
\label{eq:alpha}
    \left(r^2fgh\alpha'\right)' = \frac{k^2r^2f}{h}\sinh{4\alpha}\,.
\end{align}
If there were two horizons $g(r_h)=g(r_i)=0$ with outter horizon $r_h$ and inner horizon $r_i$,  
from \eqref{eq:alpha} we would have
\begin{align}
    0=\int_{r_i}^{r_h}dr\left(\frac{r^2fgh\alpha'}{\alpha}\right)'  = \int_{r_i}^{r_h}dr\left(\frac{k^2 r^2 f}{h}\frac{\sinh{4\alpha}}{\alpha}-\frac{r^2 fhg \alpha'^2}{\alpha^2}\right) \,.
\end{align}
The first equality is because of $g(r_h)=g(r_i)=0$. However, since we have $g(r)<0$ between the two horizons\footnote{Note that $f, h$ are  set to be positive from the boundary value, and should be always positive along the radial direction otherwise the Kretschmann scalar is divergent at the location where $f=0~\text{or}~h= 0$.} which implies the integrand of right hand side is positive. Therefore there can not be an inner horizon for the charged helical black holes we considered.

\subsection{The interior structure}
\begin{figure}[h!]
\begin{center}
\includegraphics[
width=0.43\textwidth]{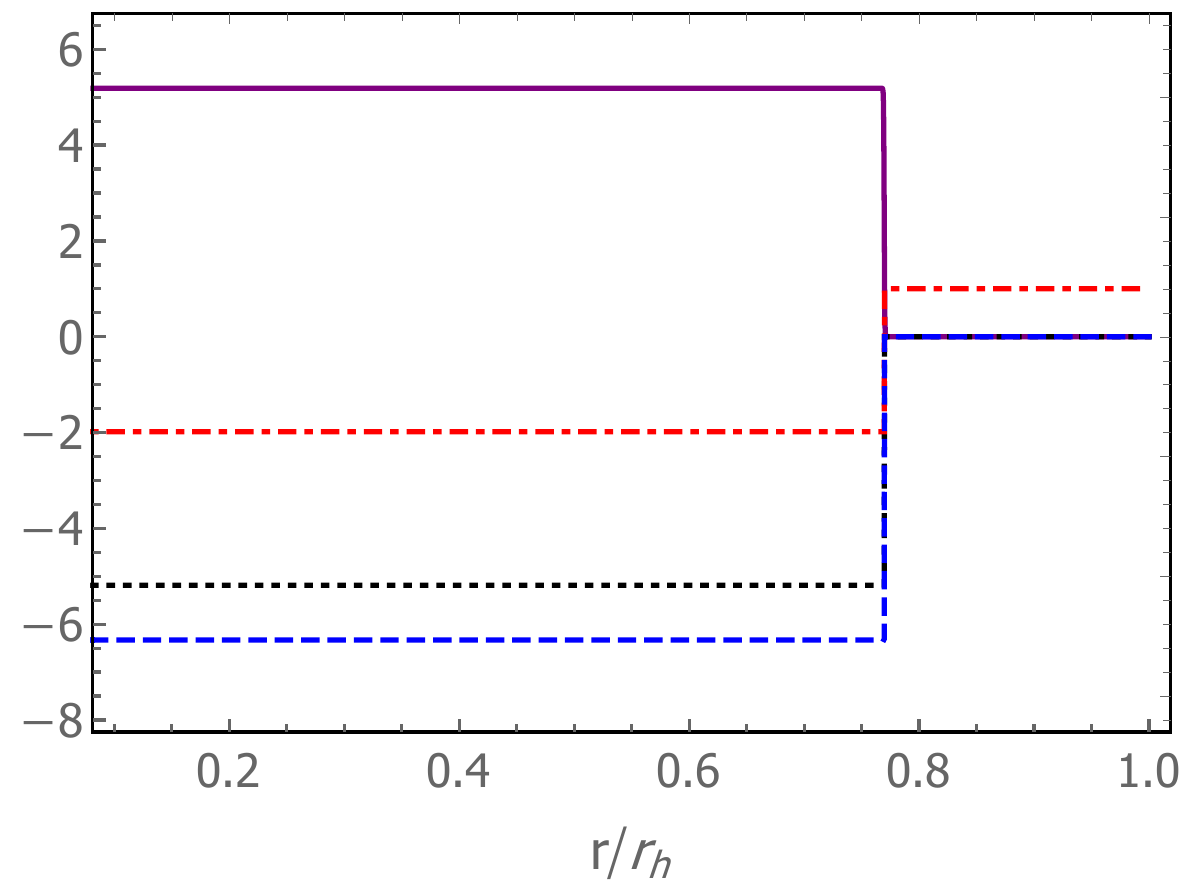}\,~~~~~~
\includegraphics[
width=0.445\textwidth]{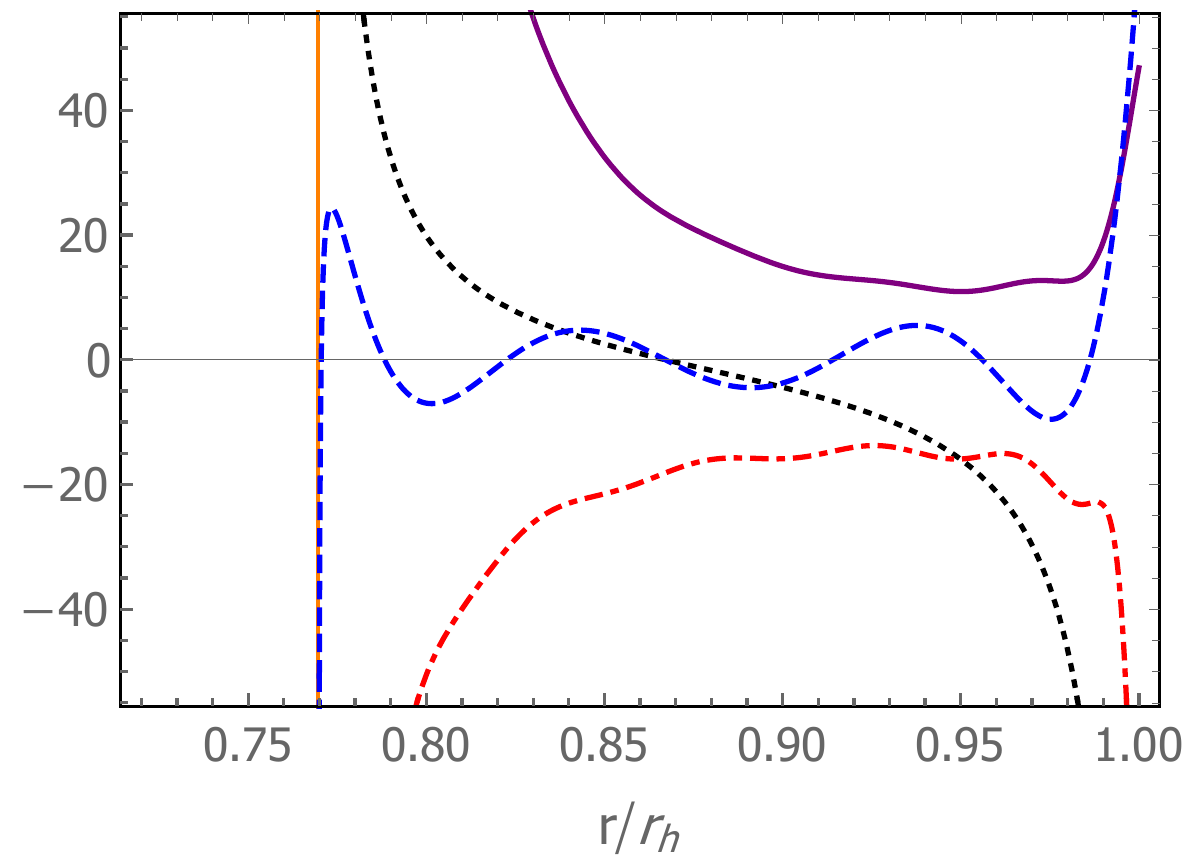}
\end{center}
\vspace{-0.7cm}
\caption{\small The interior configuration of charged helical black hole with $\alpha_b=0.01\,, ~T/\mu=0.1\,, ~k/\mu=7$. {\em Left}: The plots of $10^{-27} r f'/f$ (purple), $10^{-27} r g'/g$ (black), $r h'/h$ (red) and $10^{-12} r \alpha'$ (blue) as functions of $r/r_h$. Note that the corners of these plots are smooth. {\em Right}: The plots in the region between the event horizon and the would-be inner horizon $r/r_h=0.7696$ (the orange vertical line) of $r f'/( 10 \alpha_h^2 f)$ (purple), $r g'/g$ (black), $(r h'/h-1)/(4 \alpha_h^2)$ (red) and $r \alpha'/(4\alpha_h)$ (blue) where $\alpha_h$ is the horizon value of $\alpha$. 
} 
\label{fig:cconf}
\end{figure}

We can solve the systems of charged helical black holes numerically from the boundary to the singularity for given $\alpha_b, k/\mu, T/\mu$ and we focus on the interior structure. 
In Fig. \ref{fig:cconf} we show one example of the plot with the existence of oscillation regimes. Note that the horizon is located at $r/r_h=1$ and the singularity is at $r=0$. We find that from the horizon to the singularity, it is clear that there are three distinct dynamical regimes: including firstly the  oscillations of the metric fields, secondly the collapse of the Einstien-Rosen bridge, and finally a stable Kasner regime. In the left plot, the Kanser regime clearly exists and we have checked until $r/r_h\sim 10^{-100}$ to confirm that the Kasner geometry is stable.\footnote{We have checked that the configurations from numerics are smooth.} The right plot (with proper rescales) is an enlarged version of the oscillation regime and the collapse of the Einstein-Rosen bridge. We identify the would-be inner horizon as the location of the Reissner-Nordstr\"{o}m black hole with same $T/\mu$. For the parameter we considered here $p_t$ is close to 1 and we find that from the horizon to the would-be inner horizon, $g_{tt}$ behaves quite similar to the Reissner-Nordstr\"{o}m black hole.  Close to the would-be inner horizon, $rf'/f$ (or $g_{tt}$) has a rapid ``jump" and we call this the collapse of the Einstein-Rosen bridge following \cite{Hartnoll:2020rwq, Hartnoll:2020fhc}. Note that intriguingly the rapid  collapse of the  Einstein-Rosen bridge occurs after the oscillations in the interior time, which is 
completely different from the previous studies on the interior dynamics driven by the scalar or vector field. It should be emphasized that this behavior is quite general for any configurations with oscillation regime in the charged helical black holes.\footnote{Similar to the neutral case, the frequency of oscillations for $rf'/f$ and $rh'/h$ are roughly two times of $r\alpha'$.} Similar to the neutral helical black holes, here the spatial components of the metric $e^{2\alpha}$ and $e^{-2\alpha}$ oscillate. The oscillations of the metric field with spatial components have also been found in the example of black holes with vector hair \cite{Cai:2021obq}. Beside the relative location of oscillations and collapse in the interior time is different here, the other difference is that here the non-trivial interior structure is induced by the ``deformed" metric field $\alpha$.

\begin{figure}[h!]
\begin{center}
\includegraphics[
width=0.32\textwidth]{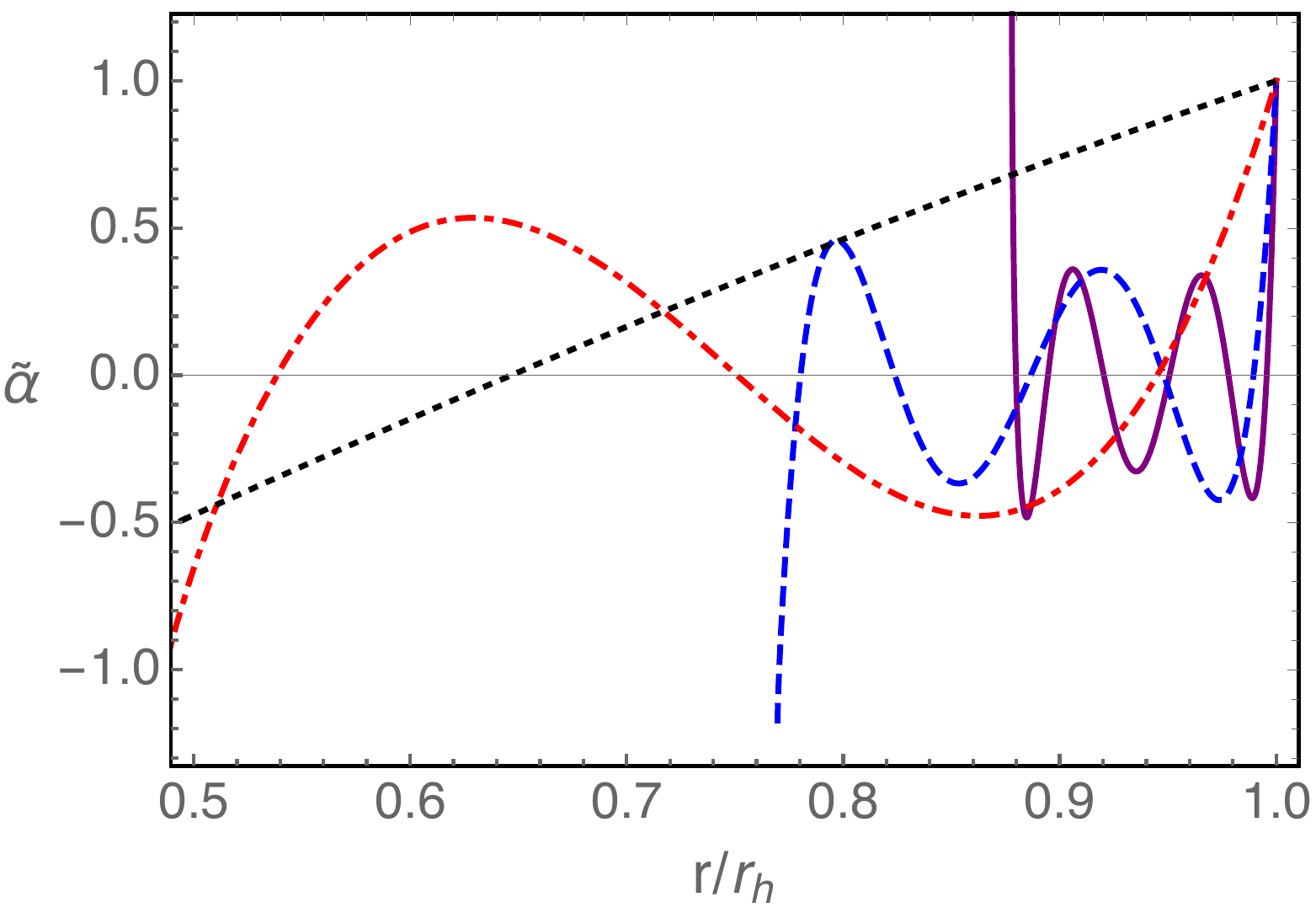}\,~~~~~~
\includegraphics[
width=0.32\textwidth]{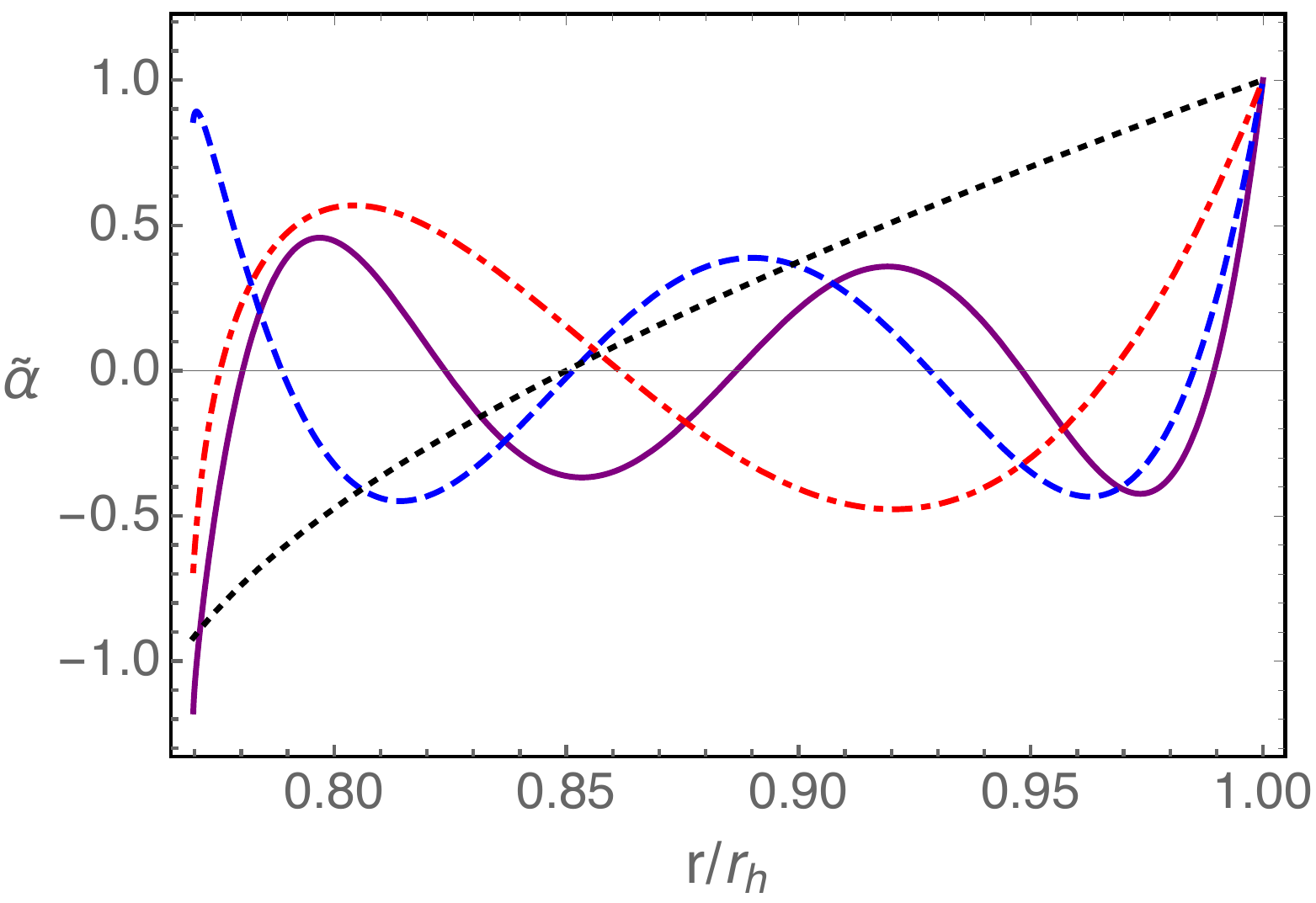}\,~~~~~~
\includegraphics[
width=0.32\textwidth]{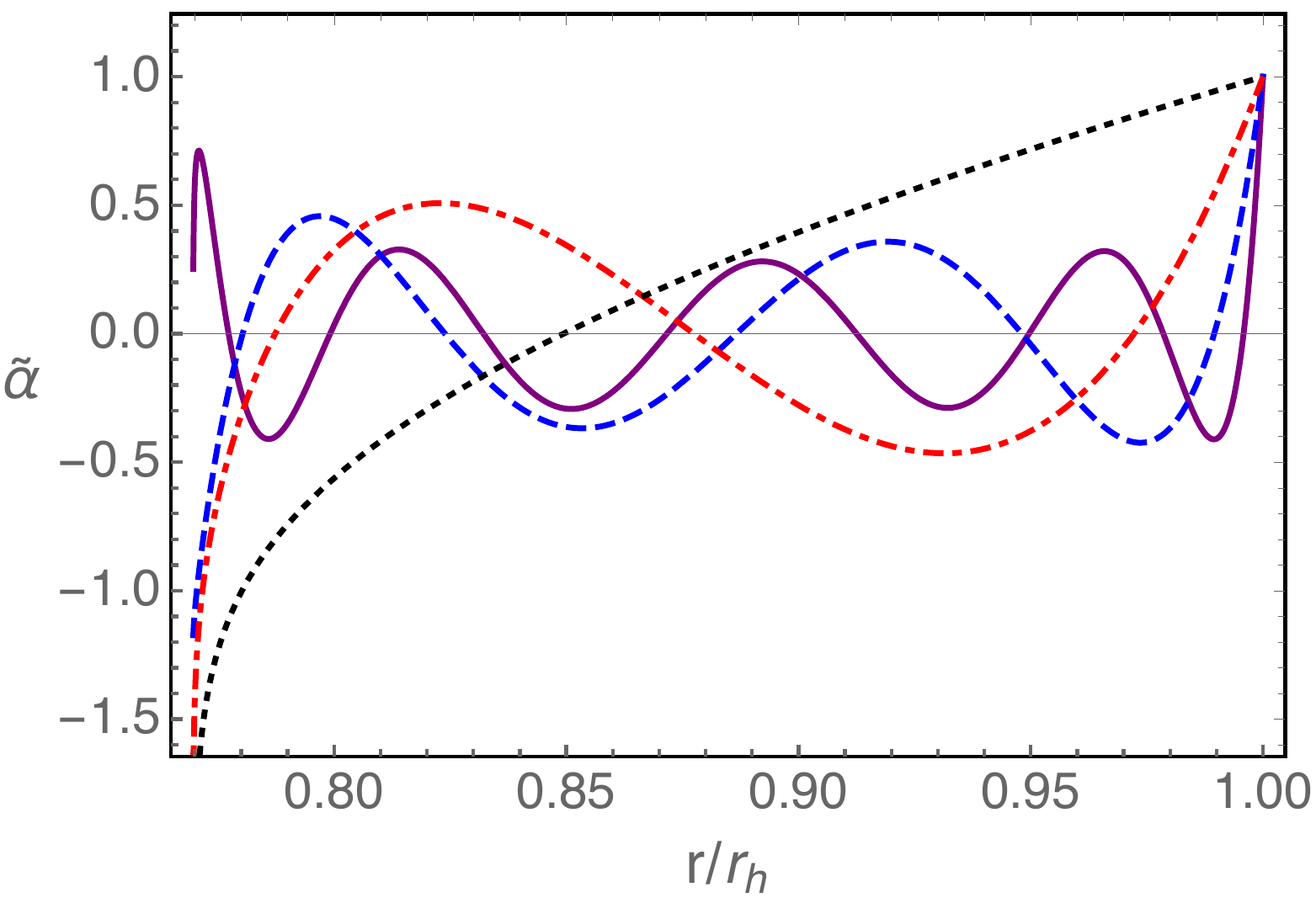}
\end{center}
\vspace{-0.7cm}
\caption{\small The plots of $\tilde \alpha=\alpha/\alpha_h$ along radial direction in the oscillation region for fixing $\alpha_b=0.01\,, k/\mu=5$ ({\em left}) while 
$T/\mu =0.05$ (purple), $0.1$ (blue), $0.3$ (red),  $1$ (black),  for fixing $T/\mu=0.1\,, k/\mu=5$ ({\em middle}) while 
$\alpha_b =0.01$ (purple), $0.5$ (blue), $1$ (red),  $2$ (black), and for fixing $\alpha_b=0.01\,, T/\mu=0.1$ ({\em right}) while $k/\mu =8$ (purple), $5$ (blue), $3$ (red), $1$ (black). Note that $\alpha_h$ is the horizon value of $\alpha$.
} 
\label{fig:calpha}
\end{figure}

In Fig. \ref{fig:calpha}, we show the oscillation of the metric field $\alpha$ along the radial direction in different situations. Here we rescale $\tilde{\alpha}=\alpha/\alpha_h$ with $\alpha_h$ the value of $\alpha$ at the horizon. In the left plot, we fix $\alpha_b=0.01, k/\mu=5$ and tune the temperature $T/\mu$ from $0.05$ to $1$, and find that qualitatively the lower the temperature is, the more times the field $\alpha$ oscillates. This behavior is the same as the neutral case which is shown in the left plot of Fig.  \ref{fig:neualpha}.  Note that for the purple and blue lines in the left figure, the system quickly enters into the Kasner regime and $\alpha$ behaves like $n_\alpha \log r$ afterwards with constant $n_\alpha$. In the middle plot, we fix $T/\mu=0.1, k/\mu=5$ and tune $\alpha_b$. 
The smaller $\alpha_b$ is, the more times  $\alpha$ oscillates, which is also similar to the neutral case. In the right plot, we fix $\alpha_b=0.01, T/\mu=0.1$ and tune the wave-number $k/\mu$. We find that the larger value of $k/\mu$, i.e. smaller pitch of the helical structure, the more times $\alpha$ oscillates.

\subsection{Kasner exponents}

Similar to the neutral helical black hole, we assume 
that when $r\to 0$, metric fields have the form 
\begin{align}
\label{eq:scalingsolc}
    \begin{split}
    g= -\frac{g_0}{r^{n_g}}+\dots\,, ~~~f= f_0r^{n_f}+\dots\,, ~~~h= h_0 r^{n_h}+\dots\,, ~~~\alpha= n_{\alpha} \log r + \alpha_0+\dots\,,
    \end{split}
\end{align}
then the gauge field $\psi$ can be obtained by integrating (\ref{eq:psi})
\begin{align}
\label{eq:sca-psi}
    \psi = \psi_i-\frac{2f_0 (n_h+2) \psi_0}{2h_0 (2n_h-n_{\alpha}^2+1)} r^{\frac{2(n_{\alpha}^2-2n_h-1)}{n_h+2}}+\dots\,,
\end{align}
where $\psi_i$ is an integration constant which is determined by the condition $\psi_h=0$.

With above ansatz, the equations \eqref{eq:ceom} can be simplified to 
\begin{align}
\label{eq:rto0c}
\begin{split}
    \frac{f'}{f}+\frac{hh'-rh'^2-2rh^2\alpha'^2}{h(2h+rh')}&=0\,,\\
    \frac{g'}{g}+\frac{2(rhh'+r^2h'^2+h^2(1+r^2\alpha'^2))}{rh(2h+rh')}&=0\,,\\
    h''-h'(\frac{h'}{h}-\frac{1}{r})&=0\,,\\
    \alpha''+\frac{\alpha'}{r}&=0\,,
\end{split}
\end{align}
which are the same as the neutral case. 
The analytical solution of \eqref{eq:rto0c} have the same form as \eqref{eq:nfngn} with  
 \begin{align} \label{eq:nfng}n_f=\frac{n_h(n_h-1)+2n_{\alpha}^2}{n_h+2}\,,~~~~~ n_g=\frac{2(n_h^2+n_h+n_{\alpha}^2+1)}{n_h+2}\,. \end{align}

We have used the assumption that the terms ignored in the above equations should be sub-leading which constrains
\begin{align}
\label{eq:assc}
\begin{split}
    n_g>-2\,,~ n_g-2n_h-2>0\,,~ n_g-2n_h+2>0\,,~ n_g-2n_h+2\pm 4n_{\alpha}>0 \,.
\end{split}
\end{align}
The second inequality, which has not presented in neutral case, comes from the terms of Maxwell field in equations \eqref{eq:ceom}. From (\ref{eq:nfng}) and (\ref{eq:assc}) we obtain 
\begin{align}
\label{eq:neqc}
n_h>-2\,, ~~~n_g> 1\,, ~~~n_{\alpha}^2>2 n_h+1\,,~~ n_\alpha^2+3\pm 2n_\alpha(n_h+2)>0\,.
\end{align}
Obviously the power exponent of \eqref{eq:sca-psi} is positive and the leading term is the constant term. 

Evaluating the conserved charge \eqref{eq:ncc} at the singularity and the horizon, we obtain
\begin{align}
    \frac{4f_0g_0h_0 (3n_h-n_{\alpha}^2+3)}{n_h+2} -4 \psi_0 \psi_i  = \frac{h_h r_h^3}{3f_h}(24f_h^2-\psi_1^2) = 2f_bh_b Ts\,,
\end{align}
where the first term should be non-negative because we set $f_b>0, h_b>0$ at boundary and the temperature and entropy density must be equal or greater than $0$. 
We have numerically check that all the above inequalities \eqref{eq:neqc} 
are satisfied and this indicates the Kasner geometry is stable. This is quite similar to the charged black holes with neutral scalar field where Kasner geometry is stable \cite{Hartnoll:2020rwq}.

Since the ansatz \eqref{eq:scalingsolc} of the solution near the singularity and the equations \eqref{eq:rto0c} are the same as neutral case, we can perform the coordinate transformation 
$\tau=\frac{2}{\sqrt{g_0}(n_g+2)}r^{(n_g+2)/2}$ to obtain the Kasner form for \eqref{eq:scalingsolc}, 
\be
ds^2= -d\tau^2+ c_t \tau^{2p_t}dt^2 + c_1 \tau^{2p_1}\omega_1^2 + c_2 \tau^{2p_2}\omega_2^2 + c_3 \tau^{2p_3}\omega_3^2 
\ee
with the Kasner exponents 
\begin{align}
    p_t=\frac{2n_f-n_g}{n_g+2}\,,~~ p_1=\frac{2n_h}{n_g+2}\,,~~p_2=\frac{2(n_{\alpha}+1)}{n_g+2}\,,~~p_3=\frac{2(1-n_{\alpha})}{n_g+2}\,.
\end{align}
From \eqref{eq:nfng}, we have  
\begin{align}
\begin{split}
    p_t=\frac{n_{\alpha}^2-2n_h-1}{n_{\alpha}^2+n_h(n_h+2)+3}&\,,~~~ p_1=\frac{n_h(n_h+2)}{n_{\alpha}^2+n_h(n_h+2)+3}\,,\\
    p_2=\frac{(n_h+2)(n_{\alpha}+1)}{n_{\alpha}^2+n_h(n_h+2)+3}&\,,~~~p_3=\frac{(n_h+2)(1-n_{\alpha})}{n_{\alpha}^2+n_h(n_h+2)+3}\,.
\end{split}
\end{align}
The same as neutral case, from the above relations we have
\begin{align}
\label{eq:ckasnerrel}
    p_t+p_1+p_2+p_3=p_t^2+p_1^2+p_2^2+p_3^2=1\,.
\end{align}
There is a constraint on $p_t$ from \eqref{eq:neqc} 
\begin{align}
    0< p_t <1\,.
\end{align}

\begin{figure}[h!]
\begin{center}
\includegraphics[
width=0.25\textwidth]{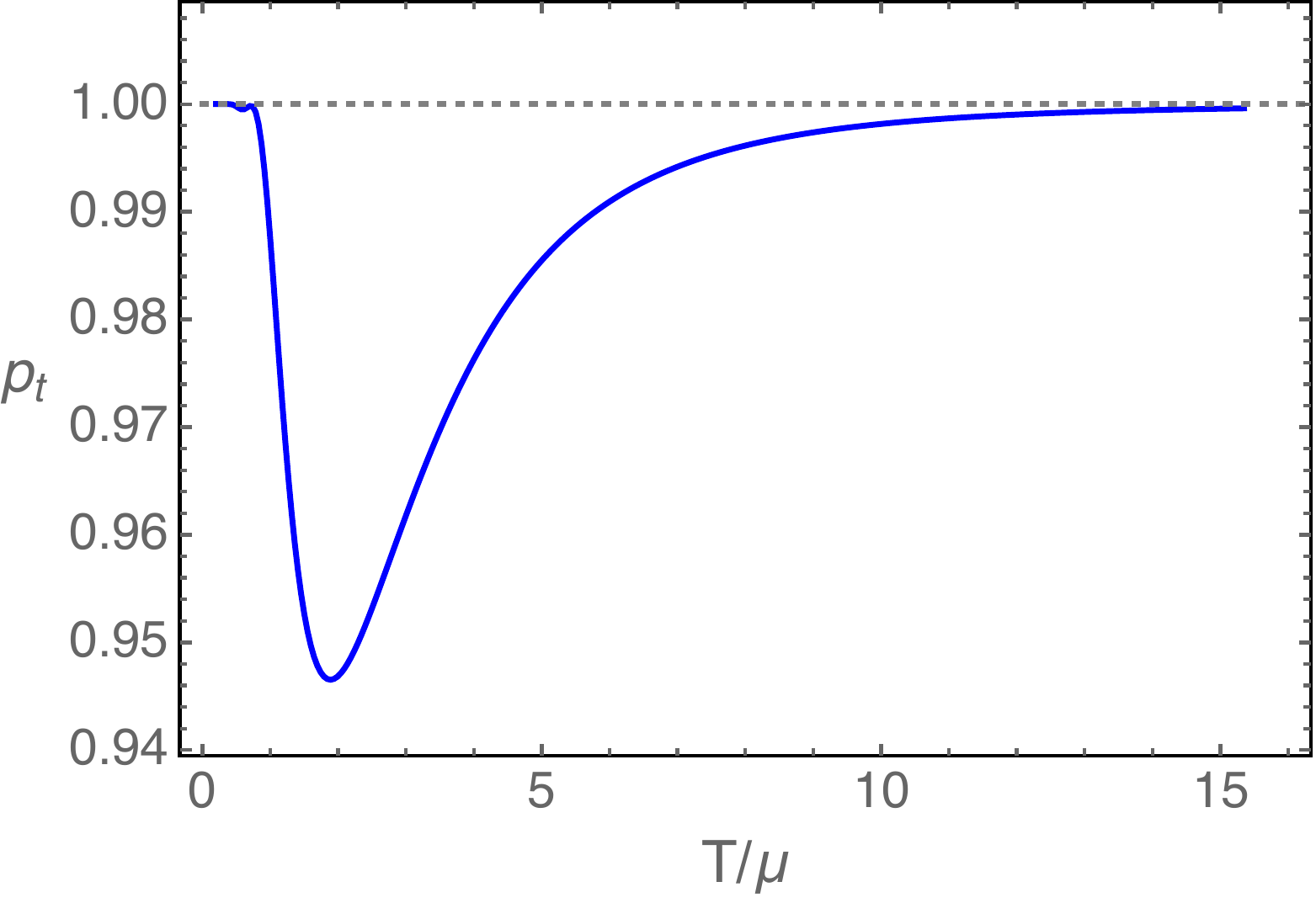}\,~~~
\includegraphics[
width=0.244\textwidth]{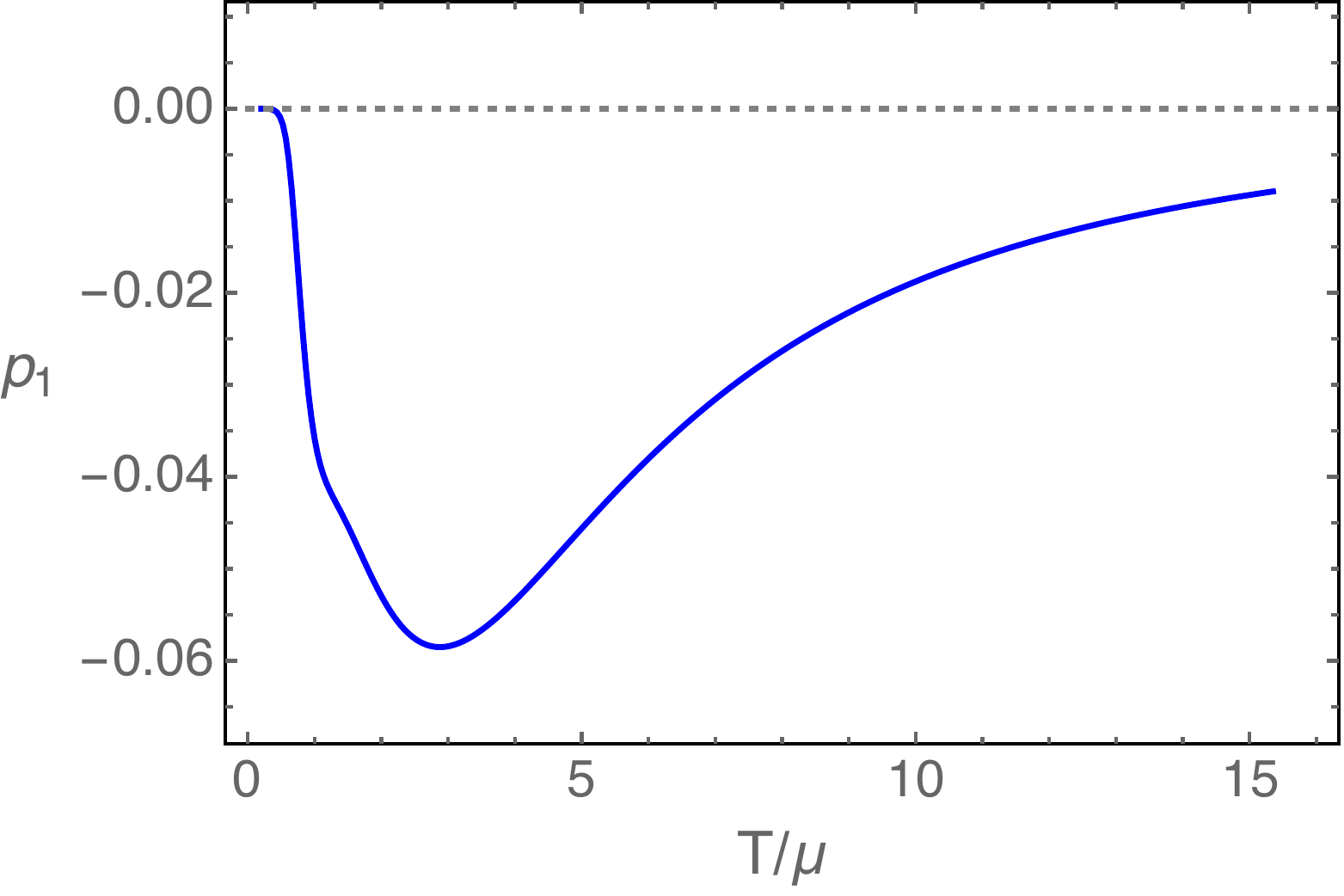}\,~~~
\includegraphics[
width=0.244\textwidth]{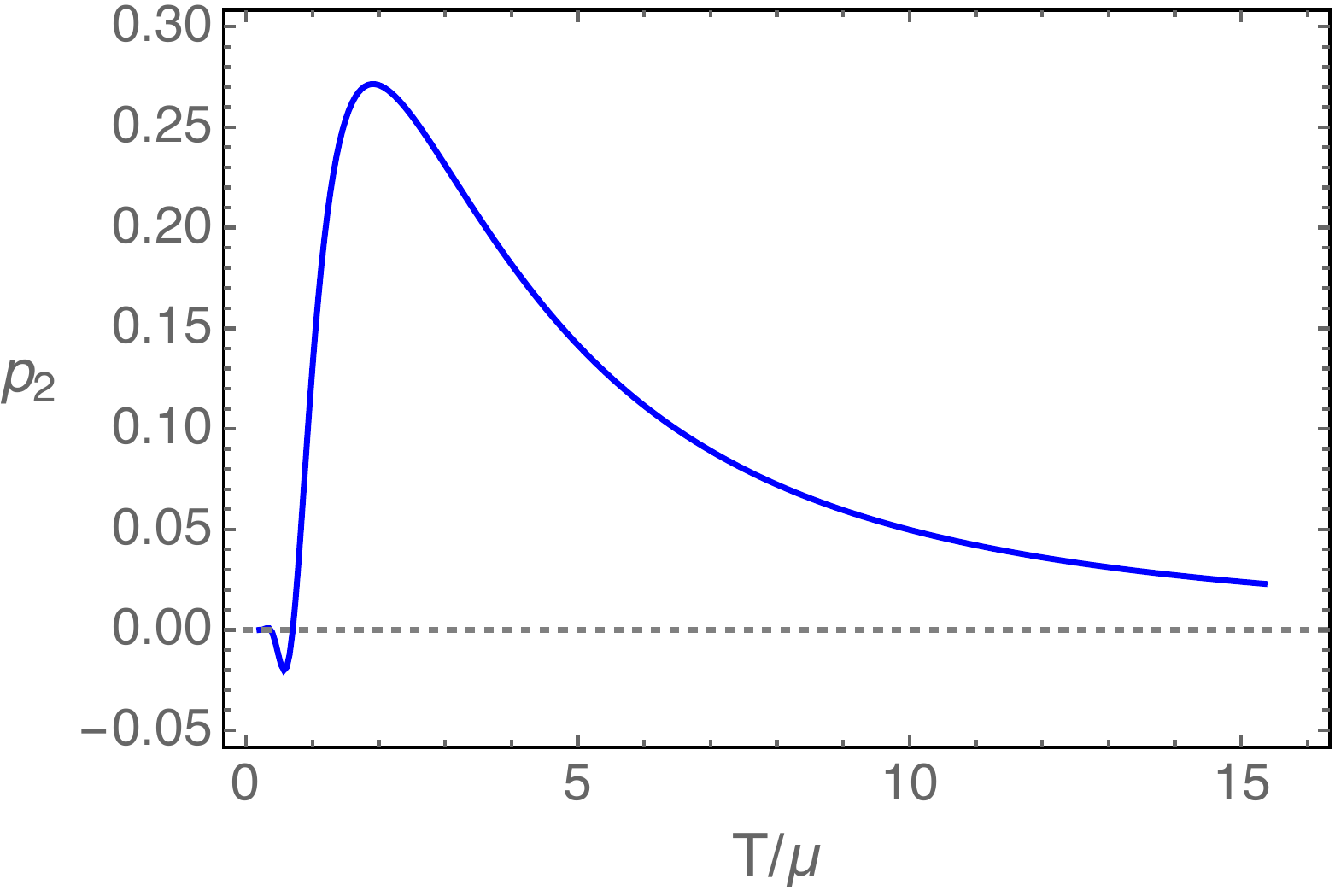}\,~~~
\includegraphics[
width=0.244\textwidth]{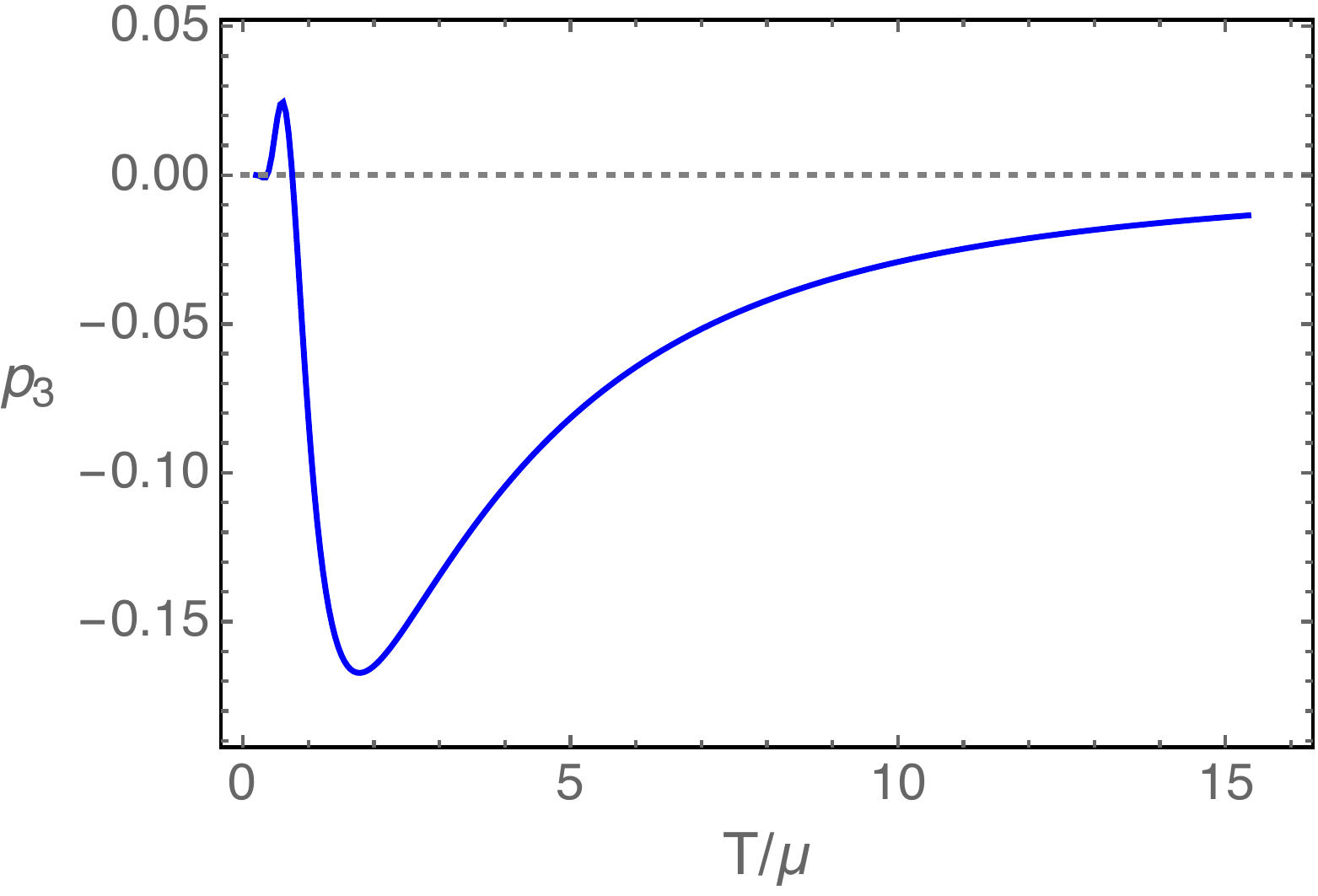}
\end{center}
\vspace{-0.7cm}
\caption{\small The Kasner exponents as functions of $T/\mu$ at $\alpha_b=0.5\,, k/{\mu}=5$. The gray dotted line is corresponding value of regular inner horizon of Reissner–Nordstr\"{o}m black hole with $p_t=1\,, p_1=p_2=p_3=0$.} 
\label{fig:cKastbm}
\end{figure}

The Kasner exponents can be obtained from the numerical solutions. In Fig. \ref{fig:cKastbm}, we show the dependence of the Kasner exponents as functions of $T/\mu$ at $\alpha_b=0.5, k/\mu=5$. One important message here is that $0<p_t<1$. When $T/\mu$ is small or very large, $p_t\to 1$. 

\begin{figure}[h!]
\begin{center}
\includegraphics[
width=0.25\textwidth]{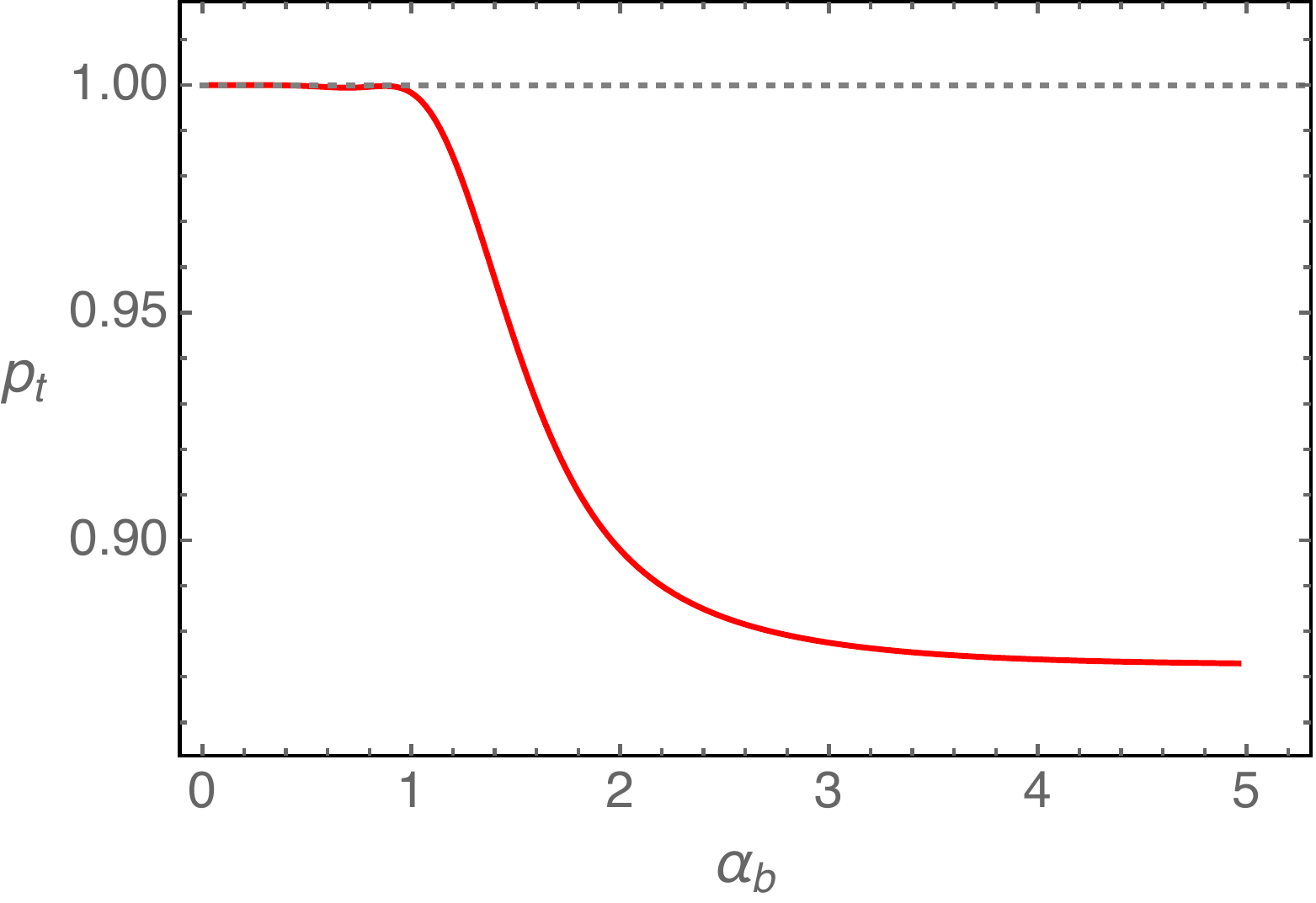}\,~~~
\includegraphics[
width=0.244\textwidth]{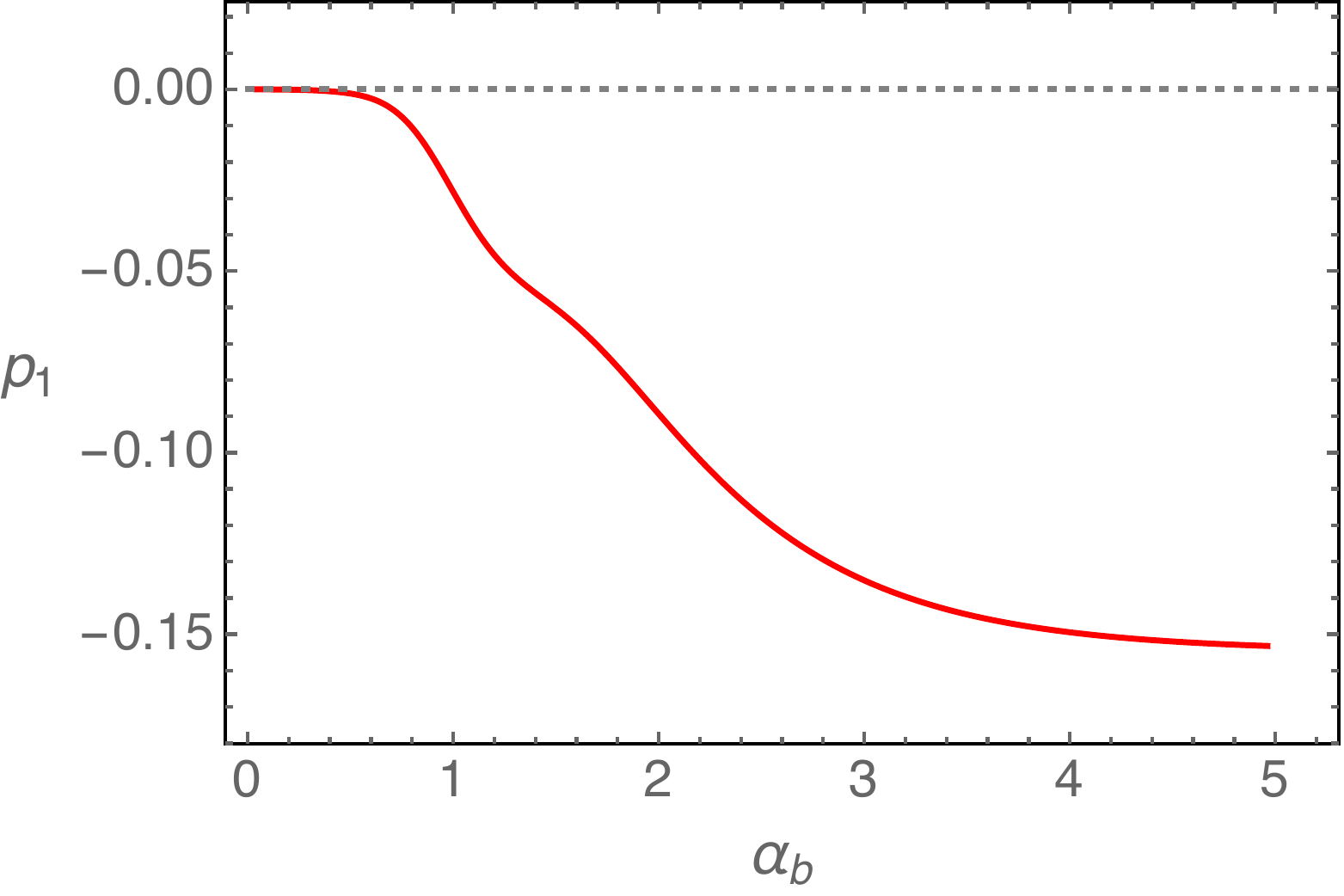}\,~~~
\includegraphics[
width=0.244\textwidth]{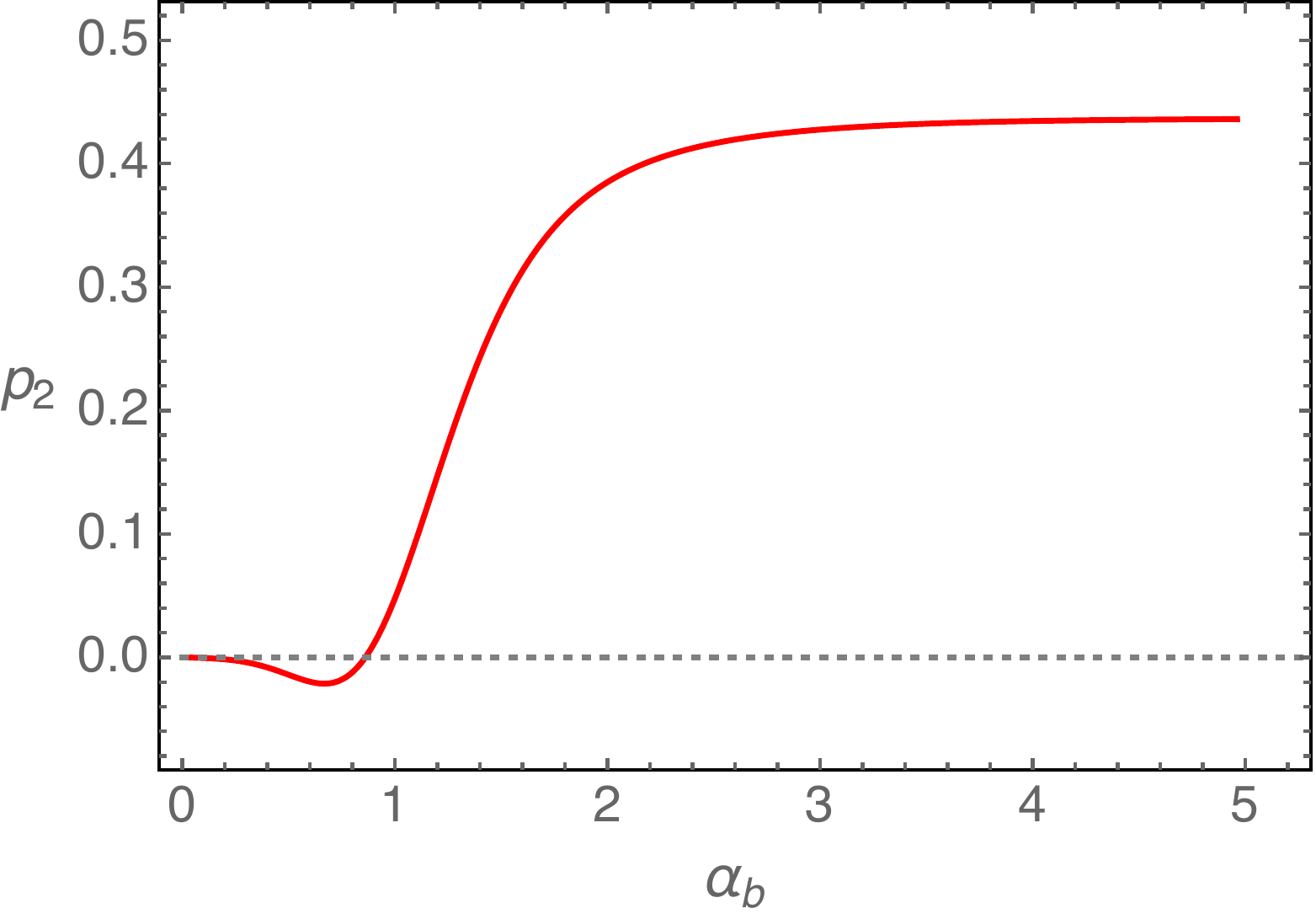}\,~~~
\includegraphics[
width=0.244\textwidth]{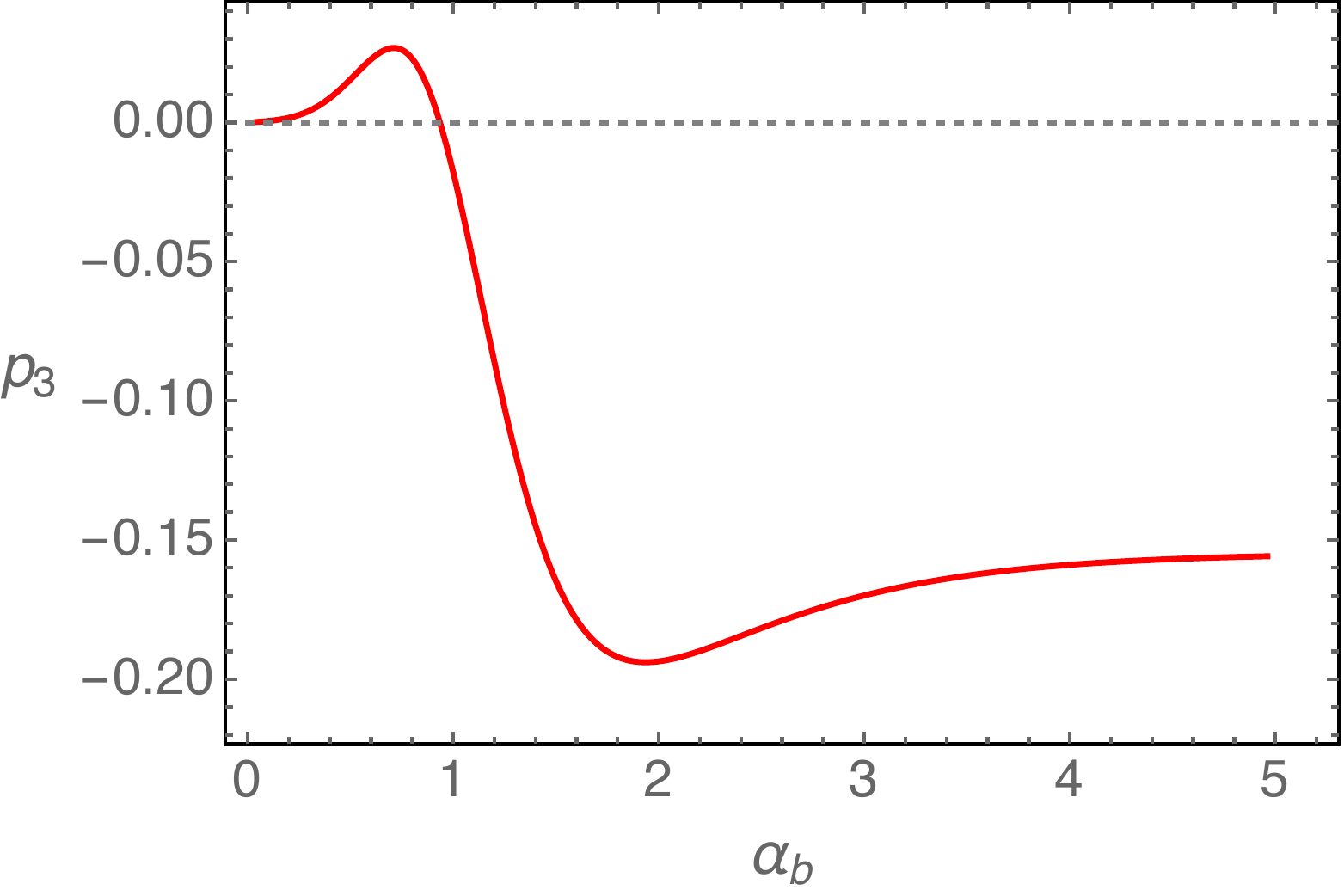}
\end{center}
\vspace{-0.7cm}
\caption{\small The Kasner exponents as functions of $\alpha_b$ at $T/{\mu}=0.5\,, k/{\mu}=5$. The gray dotted line is corresponding value of regular inner horizon of Reissner–Nordstr\"{o}m black hole.}
\label{fig:cKasalpha}
\end{figure}

In Fig. \ref{fig:cKasalpha}, we show the dependence of the Kasner exponents as functions of $\alpha_b$ at $T/\mu=0.5, k/\mu=5$. The qualitative behaviors of $-p_t, p_i,~(i=1,2,3)$ depending on $T/\mu$ or $\alpha_b$ are quite similar to $p_t, p_i,~(i=1,2,3)$ in the case of neutral helical black holes in Fig. \ref{fig:neuKasalpha}. 

\begin{figure}[h!]
\begin{center}
\includegraphics[
width=0.25\textwidth]{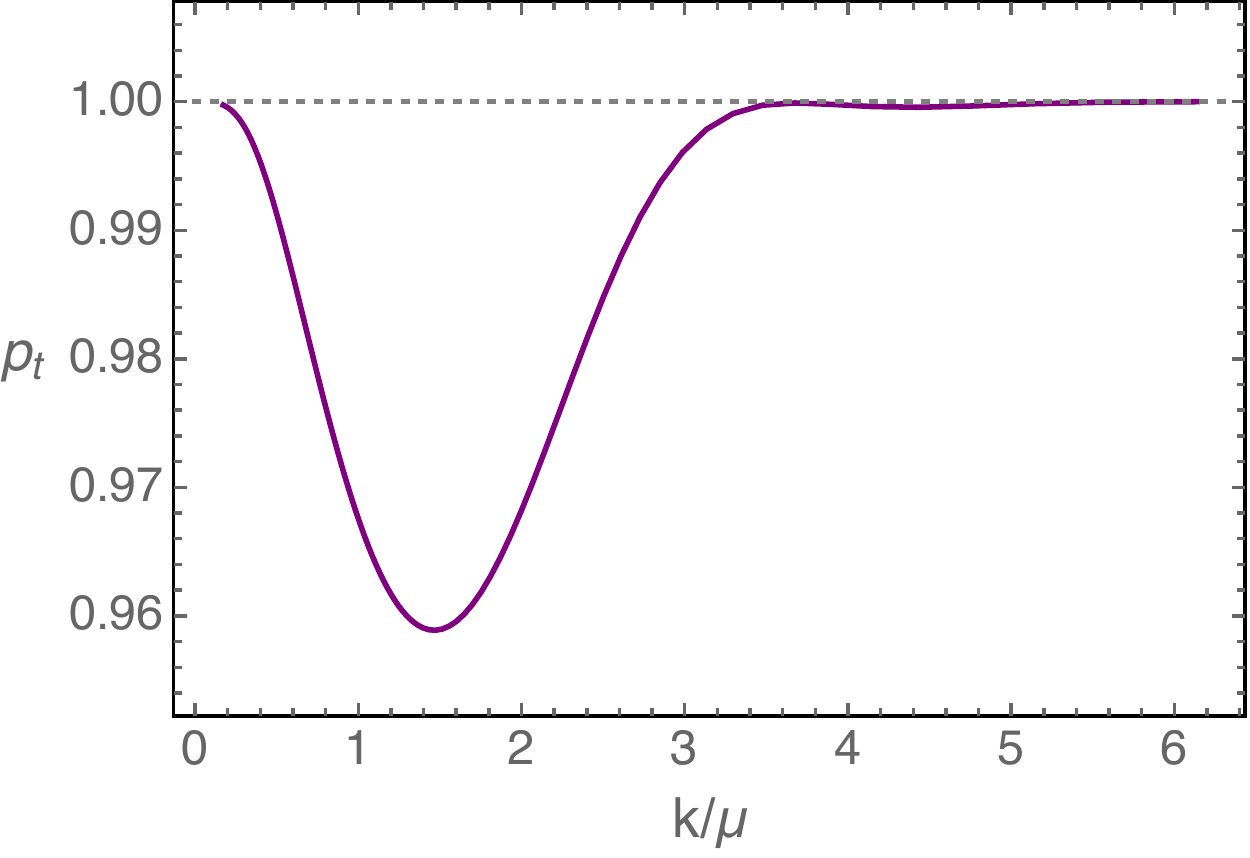}\,~~~
\includegraphics[
width=0.244\textwidth]{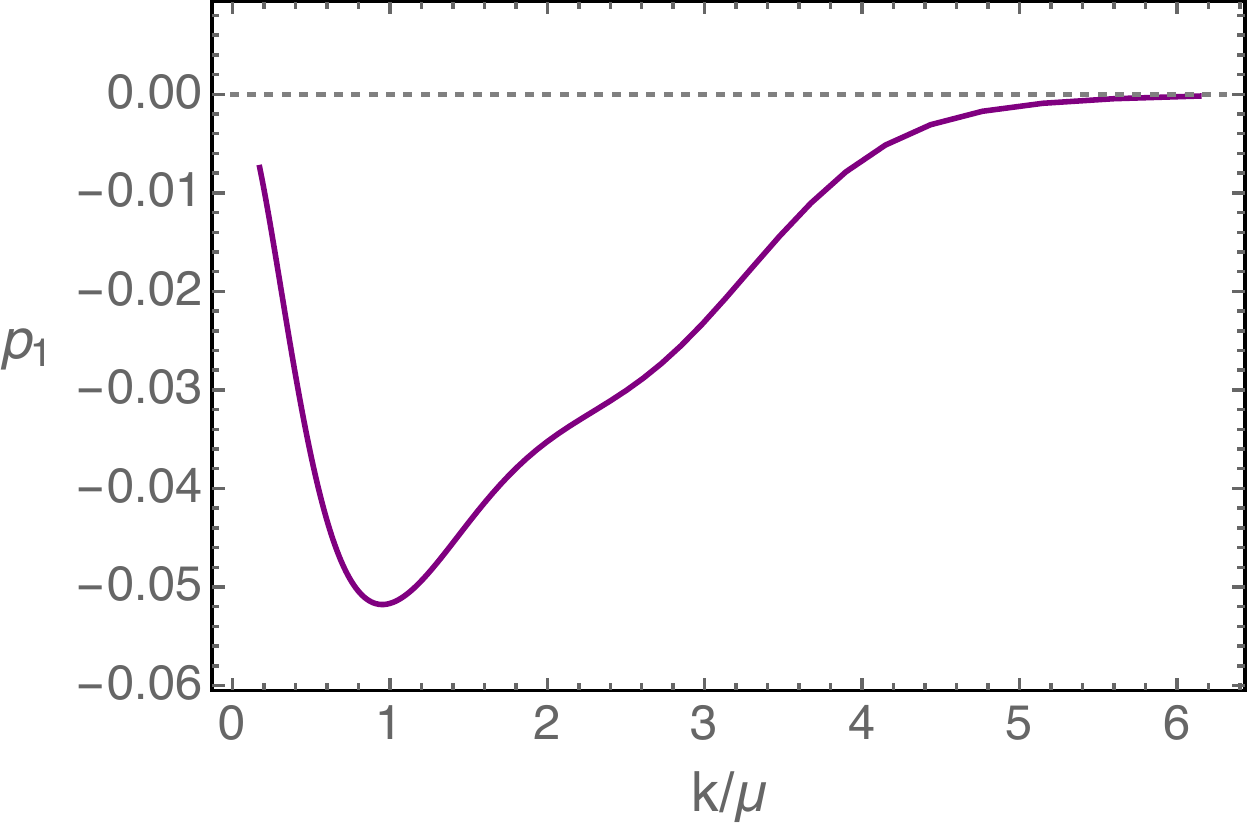}\,~~~
\includegraphics[
width=0.244\textwidth]{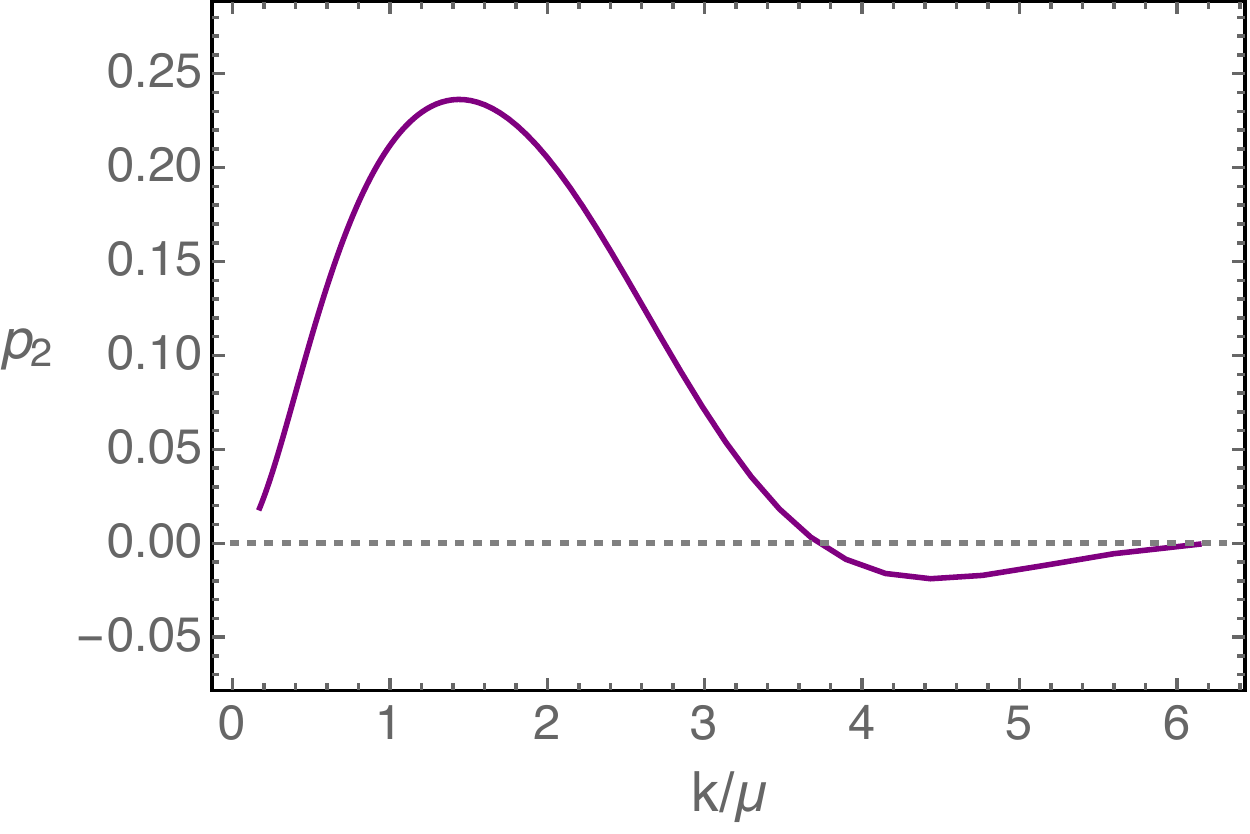}\,~~~
\includegraphics[
width=0.244\textwidth]{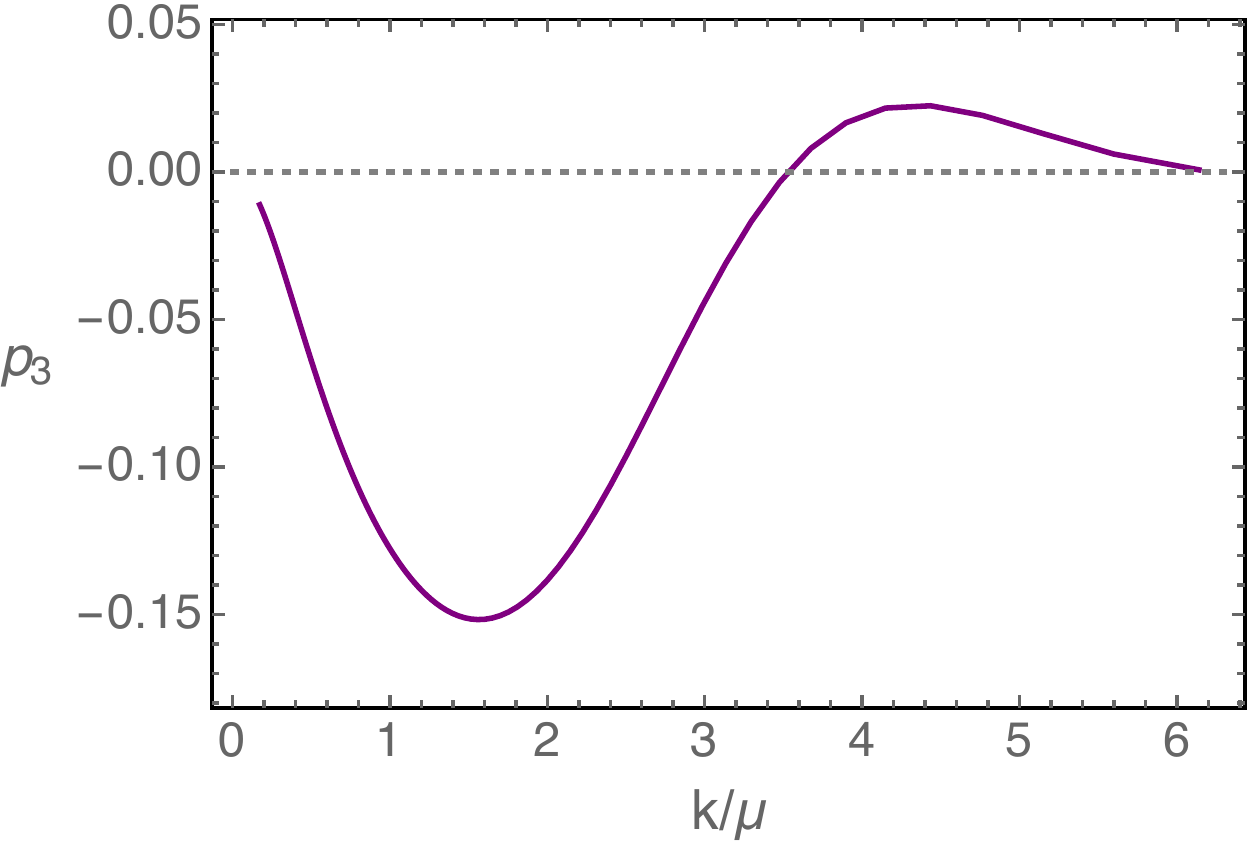}
\end{center}
\vspace{-0.7cm}
\caption{\small The Kasner exponents as functions of $k/\mu$ at $\alpha_b=0.5\,,T/{\mu}=0.5$. The gray dotted line is corresponding value of regular inner horizon of Reissner–Nordstr\"{o}m black hole.} 
\label{fig:cKaskbm}
\end{figure}

In Fig. \ref{fig:cKaskbm}, we show the dependence of the Kasner exponents as functions of $k/\mu$ at $\alpha_b=0.5\,,T/{\mu}=0.5$. In all these examples, we find that $0<p_t<1$.  We have checked that the Kasner relations \eqref{eq:ckasnerrel} are always satisfied for all the configurations we have considered.

\section{Conclusion and discussion}
\label{sec:cd}
We have studied the interior structures of neutral and charged helical black holes. One different aspect compared to previous studies in the literature is that the structure of helical black holes is induced by the deformation which is related to the metric field  while not the scalar or vector hair. For neutral helical black holes in Einstein gravity, we found that the interior geometry flows from the horizon to the spacelike Kasner singularity. At low temperature and small helical deformation strength, the metric field related to the deformation oscillates close to the horizon. This behavior is quite different from the previous examples of neutral black holes. For charged helical black holes in Einstein-Maxwell gravity, we have shown that the inner Cauchy horizon is removed by the helical deformation. The interior geometry also evolves from the horizon to a stable Kasner singularity. There is an oscillation regime at low temperatures, small helical deformation strength and small pitch of helical structure. Remarkably, the oscillation occurs near the horizon and before the collapses of the Einstein-Rosen bridge, in contrast to the known results in the literature for charged black holes with scalar or vector hair. 

For charged black holes with scalar or vector  hair, the Kasner regime might not be stable  \cite{Hartnoll:2020fhc, Cai:2021obq, Henneaux:2022ijt}. However, for charged helical black holes we have not found any example with  Kasner inversion or transition. It would be extremely interesting to study other helical black holes \cite{Donos:2012js, Donos:2012wi,  Donos:2012gg}, or to consider (neutral or charged) black holes with spatial Euclidean group broken down to symmetries of other Bianchi types different from Bianchi VII$_0$ \cite{Misner:1969hg, Iizuka:2012iv}, in order to further understand 
the dynamical chaotic behavior of Kasner geometry near the singularity for black holes induced by deformations related to the metric field.

From the holographic point of view, the interesting internal structure is expected to have corresponding descriptions in the dual field theory. Some observables  have been explored to study the physics inside the horizon, including the correlation functions \cite{Fidkowski:2003nf, Festuccia:2005pi, Grinberg:2020fdj, Frenkel:2020ysx, Liu:2021hap}, entanglement entropy \cite{Hartman:2013qma, Frenkel:2020ysx}, and complexity \cite{Caceres:2022smh, Bhattacharya:2021nqj, An:2022lvo, Auzzi:2022bfd} etc. We have shown that in the interior there are oscillation regimes for both neutral and charged helical black holes, and the collapse of the Einstein-Rosen bridge for charged helical black holes.
It was pointed out in \cite{Hartnoll:2020rwq,Hartnoll:2020fhc} that since the collapse of Einstein-Rosen bridge indicates the existence of interior extreme value of $g_{tt}$, the geodesics approximation therefore gives rise to a purely damped quasinormal mode in the dual field theory. 
When the metric field with the components of the spatial directions $\omega_2$ and $\omega_3$ oscillates, the properties of geodesics with conserved momentum along $x_2$ or $x_3$ directions might be nontrivial and subtle, thus interesting behavior might arise for the dual Green's function at large momentum. 
It would be interesting to determine the precise holographic signature for the oscillation behavior of the metric field. Finally, in the helical black holes, there are two independent Kasner exponents for the spacelike singularity. It would be quite worthwhile to explore how to characterize them using the physical quantities in the dual field theory.

\section*{Acknowledgments}
We would like to thank Ling-Long Gao for discussions.  
This work is supported by the National Natural Science Foundation of China grant No.11875083.
\appendix
\section{Expansions in neutral black holes}
\label{app:A}

Here we show the details of the expansions of fields near horizon and near boundary for the neutral helical black holes discussed in Sec. \ref{sec2}.  Close to the black hole horizon $r\to r_h$, we obtain the series solutions 
\bea
\begin{split}
    g &= 4r_h(r-r_h)+\frac{k^2 \cosh{4\alpha_h}-k^2-4h_h^2}{2h_h^2}\,(r-r_h)^2+ \cdots\,,\\
    f&= f_h\left(1 - \frac{ k^2 (-4h_h^2+k^2+3k^2\cosh{4\alpha_h})\sinh^2{2\alpha_h}}{48 h_h^4 r_h^2}\, (r-r_h)^2 + \cdots\right)\,,\\
    h&= h_h +\frac{2h_h^2-k^2\sinh^2{2\alpha_h}}{2h_h r_h}\,(r-r_h) - \frac{k^4\sinh^2{4\alpha_h}}{16 h_h^3 r_h^2}\, (r-r_h)^2+ \cdots\,,\\
    \alpha&= \alpha_h + \frac{k^2 \sinh{4\alpha_h}}{4h_h^2 r_h}\, (r-r_h) - \frac{k^2(8h_h^2+k^2-3k^2\cosh{4\alpha_h})\sinh{4\alpha_h}}{32h_h^4 r_h^2}\, (r-r_h)^2+ \cdots\,.
    \end{split}
\eea
Among five independent parameters $r_h,f_h,h_h, \alpha_h, k$ at the horizon, three of them could be removed by the scaling symmetries (\ref{eq:sym1}, \ref{eq:sym2}, \ref{eq:sym3}).  More precisely, we can use (\ref{eq:sym1}) to set $r_h=1$, (\ref{eq:sym2}) to set $f_h=1$ and use (\ref{eq:sym3}) to set $k=1$. Then we have only two free parameters $h_h, \alpha_h$ at the horizon as shooting parameters. In the dual field theory, these two parameters are related to the two scale invariant free parameters $\alpha_b, T/k$.

We get the series solution near UV  boundary $r\to \infty$ 
\begin{align}
\begin{split}
    f &= f_b \bigg(1- \frac{k^2 \sinh^2{2\alpha_b}}{6h_b^2 r^2} - \frac{c_h}{h_b r^4} - \frac{k^4 \sinh^2{2\alpha_b}}{24 h_b^4 r^4}\,(5+7\cosh{4\alpha_b})\\
    &~~~~ - \frac{\log r}{r^4} \frac{k^4}{6h_b^4}\, (\cosh{4\alpha_b}-\cosh{8\alpha_b})+ \cdots\bigg)\,,\\
    g&=r^2 \bigg( 1+\frac{k^2 \sinh^2{2\alpha_b}}{3h_b^2 r^2} + \frac{M}{r^4} + \frac{\log r}{r^4} \frac{k^4}{3h_b^4}\, (\cosh{4\alpha_b}-\cosh{8\alpha_b})+ \cdots\bigg)\,,\\
    h&= r \bigg( h_b +\frac{k^2 \sinh^2{2\alpha_b}}{2h_b^2 r^2} + \frac{c_h}{r^4} + \frac{\log r}{r^4} \frac{k^4}{6h_b^4}\, (\cosh{4\alpha_b}-\cosh{8\alpha_b})+ \cdots \bigg)\,,\\
    \alpha&= \alpha_b -\frac{k^2\sinh{4\alpha_b}}{4h_b^2 r^2} + \frac{c_{\alpha}}{r^4} + \frac{\log r}{r^4} \frac{k^4 \sinh{4\alpha_b}}{12h_b^4}\,\left( 4\cosh{4\alpha_b}-1 \right) + \cdots\,.
\end{split}
\end{align}

An asymptotic AdS black hole solution should have $f_b=1\,, h_b=1$. After integrating the EOM from horizon to boundary, we can read the value of $f\,, h$ at boundary to obtain $f_b$ and $h_b$ then we use (\ref{eq:sym2}) to set $f_b=1$ and use (\ref{eq:sym3}) to set $h_b=1$. With the data at the horizon, we can further obtain the interior configurations by integrating the system from the horizon to the singularity. Equivalently, with the boundary values of $\alpha_b$ and $T/\mu$, the whole system for the boundary to the singularity are completely determined. 

\section{Expansions in charged black holes}
\label{app:b}
In this appendix, we show the calculation details of the series solutions near the horizon and near the boundary for the charged helical black holes studied in Sec. \ref{sec3}. We obtain the IR series solutions near the  horizon $r\to r_h$
\begin{align}
\begin{split}
    g&= r_h\left(4-\frac{\psi_1^2}{6f_h^2}\right)(r-r_h)+\cdots\,,\\ 
    f&=f_h + \frac{12f_h^3 k^2 \psi_1^2 \sinh^2{2\alpha_h}}{r_h h_h^2 (\psi_1^2 -24f_h^2)^2} (r-r_h)+\cdots\,,\\
    h&= h_h+ \left(\frac{h_h}{r_h}+\frac{12f_h^2 k^2  \sinh^2{2\alpha_h}}{r_h h_h (\psi_1^2 -24f_h^2)}\right) (r-r_h) + \cdots\,,\\
    \alpha&=\alpha_h + \frac{6f_h^2 k^2 \sinh{4\alpha_h}}{r_h h_h^2 (\psi_1^2 -24f_h^2)} (r-r_h) + \cdots\,,\\
    \psi&=\psi_1(r-r_h) + \cdots\,.
    \end{split}
\end{align}
where we choose the gauge $\psi(r_h)=0$. Similar to the neutral case, we will use (\ref{eq:csym1}) to set $r_h=1$, (\ref{eq:csym2}) to set $f_h=1$ and use (\ref{eq:csym3}) to set $k=1$. Then the only three free shooting parameters at the horizon are $h_h, \alpha_h, \psi_1$, which are related  to the scale invariant parameters $T/{\mu}, k/{\mu}, \alpha_b$ in the dual field theory.

The UV expansions near the boundary $r\to \infty$ are
\begin{align}
\begin{split}
    f& = f_b\bigg( 1- \frac{k^2 \sinh^2{2\alpha_b}}{6h_b^2 r^2}-\frac{c_h}{h_b r^4}- \frac{k^4\sinh^2{2\alpha_b}}{24h_b^4 r^4}(5+7\cosh{4\alpha_b}) \\
    &~~~~~~~ -\frac{\log r}{r^4}\frac{k^4}{6h_b^4}(\cosh{4\alpha_b}-\cosh{8\alpha_b})+\cdots\bigg)\,,\\
    g& = r^2 \bigg( 1+\frac{k^2 \sinh^2{2\alpha_b}}{3h_b^2 r^2} +\frac{M}{r^4} +\frac{\log r}{r^4}\frac{k^4}{3h_b^4}(\cosh{4\alpha_b}-\cosh{8\alpha_b})+\cdots \bigg)\,,\\
    h& = r \bigg( h_b+\frac{k^2 \sinh^2{2\alpha_b}}{2h_b r^2} +\frac{c_h}{r^4} +\frac{\log r}{r^4}\frac{k^4}{6h_b^3}(\cosh{4\alpha_b}-\cosh{8\alpha_b})+\cdots \bigg)\,,\\
    \alpha & = \alpha_b -\frac{k^2 \sinh{4\alpha_b}}{4h_b^2 r^2} +\frac{c_{\alpha}}{r^4} + \frac{\log r}{r^4}\frac{k^4}{12h_b^4}\sinh{4\alpha_b}(4\cosh{4\alpha_b}-1)+\cdots\,,\\
    \psi & = \psi_b + \frac{\rho}{r^2}- \frac{k^2 \rho \sinh^2{2\alpha_b}}{3h_b^2 r^4} +\cdots\,.
    \end{split}
\end{align}
Similarly when we obtain a numerical solution, we need to use (\ref{eq:csym2}) and (\ref{eq:csym3}) to set $f_b=1$ and $h_b=1$ respectively.

\end{document}